\documentclass[preprint,12pt]{elsarticle}

\usepackage{graphicx}
\usepackage{subfigure}
\usepackage{amsmath,amssymb}

\journal{}

\newenvironment{Remark}{\vspace{0.2cm} \noindent \textbf{Remark: }}

\begin{document}

\begin{frontmatter}


\title{Analytical study on phase transition of shape memory alloy wire under uniaxial tension}
\author{Zilong Song}
\ead{buctsongzilong@163.com}
\address{Department of Mathematics and Statistics, York University, Toronto, Ontario, Canada}





\begin{abstract}
This paper considers the stress-induced phase transitions of shape memory alloy slender cylinder, and analytically studies the phase transition process and the associated instability. A three-dimensional (3D) phenomenological model with an internal variable is adopted, which is simplified to a 1D system of two strains in three regions (austenite, martensite and phase transition regions). Suitable boundary conditions and interface conditions are proposed. Theoretically, it is a free boundary problem, as the position and shape of phase interfaces are unknown. We then focus on planar interfaces (which are energetically favored), and a symmetric case when phase transition occurs in the middle. For given applied stress, two-region solutions, three-region solutions and the connecting solutions between them are obtained analytically or semi-analytically, including many period-k solutions.

Two-region solutions show that phase transition does not take place at one point, but simultaneously in a small region. The width of phase transition region is found analytically,  revealing the roles of the material and geometrical parameters. Three-region solutions represent localized inhomogeneous deformations in experiments, and capture that the stress stays almost at the Maxwell stress during propagation of transformation front. For displacement-controlled process, the transition process is demonstrated by the stress-strain curve, which captures the stress drop/rise and the transition from homogeneous deformations to period-1 localized inhomogeneous deformations. When the radius is smaller than a critical value (given by material constants), the stress drop is very sharp due to transition of solutions in a snap-back bifurcation. These features show good agreement with experimental observations and shed light on the difficulties of numerical simulations.

\end{abstract}


\end{frontmatter}



\section{Introduction}

As a smart material, shape memory alloys (SMAs) such as Ni-Ti have been utilized for various real-world applications in aerospace, automotive and biomedical industries, and oil exploration \citep{lagoudas1}. One impressive example is the Ni-Ti self-expanding stent used in the minimally invasive surgery of heart disease. Many applications are motivated by two remarkable properties of SMAs: shape memory effect and pseudo-elasticity. This paper focuses on pseudo-elasticity, which means that SMAs can undergo large recoverable strain without permanent damage during the loading-unloading cycle at relatively high temperatures \citep{huo}. The mechanism of this property is closely related to the stress-induced martensitic phase transitions \citep{shaw}.

SMAs have attracted great research interests in the past few decades, including both experiments and theoretically modelling.  The macroscopic behaviour of SMAs, including pseudo-elasticity, are characterized by stress-strain curves, and  a large number of experiments \citep{shaw,shaw1, zhang, sun3,he3} have been conducted on SMA wires at different temperatures under tension. For a class of SMAs (such as Ni-Ti), many experiments show strain softening behaviors in stress-strain (stress-elongation) curves. As a result,  phase transitions are realized by nucleation of martensite band and subsequent propagation of transformation fronts  along the wire \citep{tobushi,shaw1,sun2}, also called L\"{u}ders-like behavior. The distinction and roles of nucleation and propagation stresses have been observed in experiments and analytically studied in \cite{songdai}. At nucleation of a new phase, initial homogeneous deformations will be replaced by localized inhomogeneous deformations, where the strain varies rapidly across the transformation front or the phase transition region (PTR). Therefore, capturing the transition process and  how the geometrical and material parameters affect the process are of fundamental importance both theoretically and practically. However, systematic analytical investigations on the transition process and the associated instability under the three-dimensional (3D) framework are few, due to the complexity of the governing equations and the presence of free boundaries.

Much research effort has been devoted to the constitutive models for the phase transition of SMAs. At the microscopic level, the phase field (Ginzburg-Landau) models \citep{GL3, artemev1, artemev,levitas2} concentrate on the formation and evolution of martensitic microstructure, which are described by a set of kinetic equations for order parameters. They have shown success in dealing with many variants of martensite phase simultaneously and capturing the sharp interfaces. Nevertheless, the specimen treated in experiments are relatively large (e.g. 10-100 mm) for pseudo-elasticity, and so the size of PTR is much larger than that simulated in those models. It seems more suitable to investigate the above L\"{u}ders-like behavior at the macroscopic level. 

At the macroscopic level, the models can be put into two categories: sharp interface models and diffuse interface models. Some thermoelastic models \cite{abeyaratne1, abeyaratne2, abeyaratne, knowles} treat the phase interface as a sharp singular surface, where further conditions such as nucleation criterion and kinetic relation are proposed. For simulations with moving phase interface, see also \cite{Berezovski2}. In diffuse interface models, the transformation front is often identified with a small PTR, which smoothly connects the low-strain austenite phase and high-strain martensite phase with large gradients. Some models \citep{shaw4,chang,he4,he1,yu2018} include extra gradient terms (e.g., gradient of strain), whose coefficient sets a characteristic length scale for the width of PTR. One disadvantage might be that it is not an easy task to determine the artificial coefficient beforehand. The phenomenological models \citep{rajagopal, idesman,zaki} introduce the martensite volume fraction as an internal variable into the Helmholtz free energy together with a dissipation function, then both the strain and martensite volume fraction are continuous across PTR. 

On the continuum level, the transition from homogeneous deformations to localized inhomogeneous deformations is due to the material instability. This is reflected by the existence of multiple solutions at certain stress or elongation, leading to the transition to a more stable solution by the energy criterion. The current work adopts phenomenological model by \citep{rajagopal}. One advantage of such a model is that all the parameters have clear physical meaning (no artificial coefficient) and the influence of each parameter on the transition process can then be easily identified.

Many analytical results in the literature are based on 1D models without strain softening behaviour, and investigated stress-strain-temperature response of various structures (e.g., wires \citep{brinson,marfia}, layer structures \citep{wu1} and actuators \citep{kosel,shaw5,kim2017}). However, the high-dimensional effect and strain softening behaviour are crucial to capturing the stress drop and determining the width of PTR during phase transition, therefore it is desirable to take these into account. Under 3D frameworks, some analytical results \cite{liu2014}  are obtained for thick-walled cylinder subjected to external pressure, but with Tresca transformation criterion and linear hardening. For an infinitely-long cylinder, some analytical works \cite{zhong1, zhong2} treat the martensite band as inclusion in a 3D inclusion-matrix system with planar sharp interfaces, capturing some important features such as peak and propagation stresses.

Another class of analytical study was first attempted in \citep{cai,cai1} by  adopting a non-convex hyperelastic 3D energy and reducing it to a 1D rod model. It has been generalized to study the geometric size effect and hysteresis loop in  \citep{wangjiong3,wangjiong5} for SMA wires. By combining with the model in  \cite{rajagopal}, it has been extended to study SMA layer \citep{wangjiong1, wangjiong2} and geometrically graded SMA strip \citep{wangjiong2017a,wangjiong2017b}. Our previous work \citep{songdai} has studied piecewise homogeneous deformations of an SMA wire, and has obtained analytical results for peak and propagation stresses \citep{Ericksen}. Subsequently for a 3D infinitely long cylinder \citep{songdai2015}, we have studied localized inhomogeneous deformations by keeping high-dimensional effects, capturing the propagation stress plateau. The present work aims to investigate the phase transition of an SMA wire with finite length, which is a natural continuation of infinite case \citep{songdai2015}. This is not an easy task, since one has to deal with free boundaries in response to applied stress or total elongation. In this work, we aim to systematically study the transition process from the homogeneous deformation to the localized inhomogeneous deformation, and to reveal the roles of geometric and material parameters in this process.

We start from a 3D formulation \cite{rajagopal}, and the mechanical system is simplified to a 1D system for three different regions, i.e. austenite region (AR), martensite region (MR), and PTR, by utilizing coupled series-asymptotic technique \citep{dai, dai1}. Since the shape and positions of interfaces between different regions are not known, mathematically this is a tricky free boundary problem. We concentrate on the planar interfaces, which are energetically favored (see \citep{songdai2015}) and make analytical results available. Then we consider the symmetric case, where phase transition takes place in the middle of the cylinder. First, two-region solutions with AR and PTR are obtained analytically, with many period-k solutions. Second, for three-region solutions with AR, PTR and MR, determining the solutions is reduced to a system of two nonlinear algebraic equations. The case of infinitely long cylinder in \citep{songdai2015} can be recovered as a special case of such solutions. Third, connecting solutions are constructed for the gap between two-region and three-region solutions, which is due to the interaction between phase interface and middle (boundary) surface. Finally, the stress-elongation curves are obtained based on these solutions, and the transition process is captured by energy criterion.

Finally, we briefly mention the key findings of the analytical study. First, analytical solutions show that phase transition does not take place at one point, but simultaneously in a small region. The analytical formulas for the width of PTR clear reveal the roles of the material and geometrical parameters. Second, three-region solutions capture the localized inhomogeneous deformations, indicating that the stress always stays close to Maxwell stress in the stress-elongation curve. Third, the transition from homogeneous deformations to localized inhomogeneous deformations is clearly demonstrated at nucleation of a new phase. The stress drop or rise is captured in stress-elongation curves, and the roles of parameters are disclosed. For instance, as the radius of cylinder becomes smaller than a critical value (given by material constants),  there is a sharp stress drop as observed in experiments.  As multiple solutions exist for certain fixed total elongation, it provides insights into the difficulties of direct numerical computations \citep{mirzaeifar1}, e.g., mesh sensitivity and convergence difficulty. 

The manuscript is arranged as follows. Section 2 briefly recalls 3D formulation and its simplification to a 1D system in three different regions (AR, MR and PTR). 
In Section 3, a symmetric case for a finite cylinder is considered. In Sections 4 to 6, the two-region solutions, three-region solutions and connecting solutions are constructed, either analytically or semi-analytically.  Then in Section 7, the phase transition process is investigated through stress-elongation curves in displacement-controlled process. Finally, concluding remarks are made.

\section{Problem formulation and simplification}

As this work is a continuation of previous work in \citep{songdai2015}, for completeness this section is a brief re-account of the problem setup in \citep{songdai2015}, with necessary modifications and remarks. The previous work considers the case of an infinitely long cylinder, while this work studies the case of finite cylinder. Thus, the boundary conditions are different, and the aim now is to study the transition process from homogeneous deformations to localized inhomogeneous deformations and the associated instability.  Another complication in current work is that during the transition process the number and location of interfaces (regions) are not known beforehand, so it is a free boundary problem. As we will see later that, the infinitely long case \citep{songdai2015}  can be considered as a special (and idealized) case of current one for the three-region localized inhomogeneous deformations.

\subsection{Constitutive model}

This subsection briefly recalls the constitutive model proposed by Rajagopal and Srinivasa in \cite{rajagopal1, rajagopal}, used in previous analytical works \citep{wangjiong1,songdai2015}. The model is a phenomenological model with an internal variable defined as the volume fraction $\alpha$ of martensite phase. It is described by two independent functions: Helmholtz free energy $\Phi(\mathbf{F}, \alpha,T)$ and rate of mechanical dissipation $\xi(\alpha)$, where $\mathbf{F}$ is deformation gradient and $T$ is the absolute temperature. 

This work focuses on the isothermal case. By the first and second laws of thermodynamics, we have the following relations 
\begin{equation}
\label{eq1} \mathbf{\Sigma}=\frac{\partial \Phi(\mathbf{F},
\alpha,T)}{\partial \mathbf{F}} ,\quad \xi=-\frac{\partial
\Phi(\mathbf{F},\alpha,T)}{\partial \alpha}\dot{\alpha},
\end{equation}
where $\mathbf{\Sigma}$ is the nominal stress tensor (or the transpose of first Piola-Kirchhoff stress). The specification of $\Phi$ relies on the natural configuration associated with phase state $\alpha$, defined through deformation gradient tensor 
\begin{equation}
\label{eq3}
\begin{aligned}
& \mathbf{G}_\alpha:=(1-\alpha)\mathbf{I}+\alpha \mathbf{G}, \quad 0\le \alpha \le 1,\\
&\mathbf{G}=\mathrm{diag}(1-s_1,1-s_1,1+s_2),
\end{aligned}
\end{equation}
where diag is a diagonal matrix in cylindrical polar coordinates, $\mathbf{G}$ is the natural configuration of martensite phase, and $s_1$ and $s_2$ (usually $\ll 1$) are two material parameters, called the specific strains normal to and along the axis of the cylinder. The elastic part of the deformation gradient $\mathbf{F}$ is $\mathbf{F}_\alpha=\mathbf{F} \mathbf{G}_\alpha^{-1}$, and the Helmholtz free energy per reference volume is given by
\begin{equation}
\label{eq5}
\begin{aligned}
\Phi(\mathbf{F},\alpha,T)= & \det\mathbf{G}_\alpha
[(1-\alpha)\Phi_1(\mathbf{F}_\alpha)+\alpha\Phi_2(\mathbf{F}_\alpha)]\\
& +B\alpha(1-\alpha) + [(1-\alpha)\phi_1(T)+\alpha\phi_2(T)],
\end{aligned}
\end{equation}
where $\Phi_1$ and $\Phi_2$ are respectively the strain energy
functions of the austenite and martensite phases, $B$ is interfacial constant \cite{muller}, and $\phi_1$ and $\phi_2$ are respectively the thermal free energies of the austenite and martensite phases. 

The dissipation function $\xi$ is adopted as 
\begin{equation}
\label{eq7} \xi=\begin{cases} A^+(\alpha)|\dot{\alpha}| = (k^+\alpha+Y^+ ) \dot{\alpha} ~~~~ &
\mathtt{if} \quad \dot{\alpha}\geq0,\\
A^-(\alpha)|\dot{\alpha}| = - (k^-(1-\alpha)+Y^{-} ) \dot{\alpha}~~~~ & \mathtt{if} \quad \dot{\alpha}\leq0,
\end{cases}
\end{equation}
where $\pm$ correspond to loading and unloading processes respectively,  and $k^{\pm}$ and $Y^{\pm}$ are referred to as the dissipative constants. Substituting (\ref{eq7}) into $(\ref{eq1})_2$ leads to the evolution law of $\alpha$
\begin{equation}
\label{eq10}
\begin{aligned}
-A^-(\alpha)<-\frac{\partial \Phi}{\partial \alpha}<A^+(\alpha)\quad\Rightarrow \dot{\alpha}=0,\\
\dot{\alpha}\neq0 \Rightarrow -\frac{\partial \Phi}{\partial \alpha}=\begin{cases} A^+(\alpha) \quad
& \mathtt{if} \quad \dot{\alpha}>0,\\
-A^-(\alpha) \quad & \mathtt{if} \quad \dot{\alpha}<0. \end{cases}
\end{aligned}
\end{equation}
Here $\dot{\alpha}=0$ means that the material response is elastic, e.g., when the material initially undergoes elastic deformation in pure austenite phase ($\alpha =0$). When phase transition occurs, the variable $\alpha$ in PTR is determined by $(\ref{eq10})_2$.

\Remark The interfacial energy $B\alpha (1-\alpha)$ in (\ref{eq5}) leads to a non-convex energy and is closely related to the strain-softening behaviour and the associated instability. Similar terms of $\alpha$ were introduced in other works \citep[e.g.,][]{sun,zaki} to account for mixing effect, resulting in softening response of the material. The terms in dissipation (\ref{eq7}) will also affect the strain-softening, and the constants $k^\pm$ have competing effects with $B$.

\subsection{The 3D mechanical system}

\begin{figure}
\centerline{\includegraphics[width=3in]{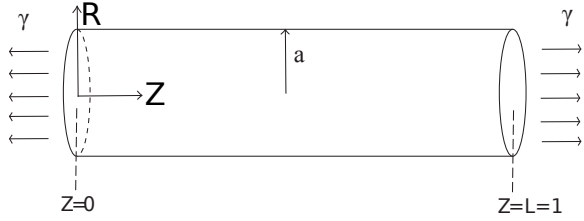}} \caption{The
cylinder in the reference configuration subjected to a
uniaxial stress $\gamma$.} \label{fig0}
\end{figure}

Figure \ref{fig0} shows the slender cylinder (wire) with finite length under uniaxial average stress $\gamma$. Without loss of generality, the length of cylinder is set to be $L=1$, so the dimensionless radius $a$ here is interpreted as ratio of radius to length, and all the displacements below are already scaled by $L$. For a slender cylinder, we have $a\ll1$. The reference configuration is entirely made up of austenite phase. The cylindrical polar coordinates are adopted, and $(R,\Theta,Z)$ and $(r,\theta,z)$ denote a generic point in the reference and current configurations, respectively. The general deformation can be described by
\begin{equation}
\label{eq11} r=R+V(R,Z),\quad \theta=\Theta,\quad z=Z+W(R,Z),
\end{equation}
where $V$ and $W$ are radial and axial displacements. The deformation gradient $\mathbf{F}$ is easily calculated by taking the derivatives, and the stress tensor $\mathbf{\Sigma}$ can be obtained from $(\ref{eq1})_1$. 

We consider a quasi-static process for the phase transition of SMAs, as in some displacement-controlled experiments.  By neglecting the body force, the mechanical field equations are
\begin{equation}
\label{eq_17} \frac{\partial \Sigma_{Rr}}{\partial R}+\frac{\partial
\Sigma_{Zr}}{\partial Z}+\frac{\Sigma_{Rr}-\Sigma_{\Theta
\theta}}{R}=0, \quad \frac{\partial \Sigma_{Rz}}{\partial
R}+\frac{\partial \Sigma_{Zz}}{\partial Z}+\frac{\Sigma_{Rz}}{R}=0,
\end{equation}
where subscripts denote the components of stress along $R,\theta,Z$ directions.
Next we consider the boundary conditions, which are different from \citep{songdai2015}. First, the traction-free boundary conditions on the lateral surface are
\begin{equation}
\label{eq_18} \Sigma_{Rr}|_{R=a}=0,\quad \Sigma_{Rz}|_{R=a}=0.
\end{equation}
Second, on the end surfaces of the cylinder, we have
\begin{equation}
\label{eq_19}
\Sigma_{Zz}|_{Z=0,1}= \gamma, \quad \Sigma_{Zr}|_{Z=0,1}= 0,
\end{equation}
where $\gamma$ is the average stress. For a force-controlled process, $\gamma$ is a given parameter, whereas for a displacement-controlled process as in many experiments, the parameter $\gamma$ is to be determined by the total displacement of cylinder.

The 3D mechanical system (\ref{eq_17}-\ref{eq_19}) for the two displacements $(V,W)$ will be coupled with the phase state variable $\alpha$, since $\mathbf{\Sigma}$ involves $\alpha$. The extra phase state $\alpha$ is determined by the evolution law $(\ref{eq10})$. 

\subsection{The 1D governing equations}

This subsection presents the simplified 1D equations in different regions based on the preceding 3D system. The derivation is based on the series-asymptotic technique in \citep{cai,cai1}, which takes advantage of two small parameters: the dimensionless radius $a$ and the scale of typical axial strain. The details are given in \cite{songdai2015}, and here we present a simple modified version of the derivation and final system, with necessary important remarks.

First, we define the radial strain by $\tilde{V} = V/R$, and take the series expansion for the two strains
\begin{equation}
\begin{aligned}
\label{eq17}\tilde{V}  (R,Z)=\sum_{k=0}^\infty \frac{\tilde{V}_{k}(Z) R^{2k}}{2(1+k)},
\quad W_Z(R,Z)=\sum_{k=0}^\infty W_{kZ}(Z) R^{2k},
\end{aligned}
\end{equation}
where and hereafter subscript $Z$ denotes derivative with respect to $Z$. In the derivation, we only keep the leading order terms of the typical axial strain. Recursive relations can be derived from (\ref{eq_17}) for high-order coefficients (terms with $k\ge 1$) in (\ref{eq17}) in terms of leading-order coefficient. They are different in three different regions: austenite region (AR, $\alpha=0$), martensite region (MR, $\alpha=1$) and phase transition region (PTR, $0<\alpha<1$). For instance, for PTR in loading process, $\alpha$ is first determined by $-\partial \Phi/\partial \alpha = A^+$, and then it is substituted into (\ref{eq_17}) to derive such recursive relations.

\Remark The series expansion (\ref{eq17}) in terms of $R^2$ is due to the symmetry of the problem, since the terms with odd powers naturally disappear. The factor $2(1+k)$ in $(\ref{eq17})_1$ is to make all recursive relations in simple and compact form  \citep{songdai2015}, otherwise the expressions of recursive relations are more  complicated (see, e.g., \cite{bostrom}).

With recursive relations, the conditions in (\ref{eq_18})  provide the 1D governing equations for the leading-order strain vector (axial and radial strains)
\begin{equation}
\begin{aligned}
\label{eq23_2}
\mathbf{U}_0 (Z)= [W_{0Z}(Z), \tilde{V}_0(Z)]^T,
\end{aligned}
\end{equation}
where superscript $T$ means transpose of a vector. By truncating the system and keeping upto $O(a^2)$ terms, the governing equations for $\mathbf{U}_0$ in three regions take the form
\begin{equation}
\begin{aligned}
\label{eq23}
& a^2 \mathbf{U}_{0ZZ} + \mathbf{H} \mathbf{U}_0 = \mathbf{f}_0, \quad \mathrm{in} \quad \mathrm{AR},\\
& a^2 \mathbf{U}_{0ZZ} + \bar{\mathbf{H}} \mathbf{U}_0 = \bar{\mathbf{f}}, ~\quad \mathrm{in} \quad \mathrm{PTR},\\
& a^2 \mathbf{U}_{0ZZ} + \mathbf{H} \mathbf{U}_0 = \mathbf{f}_1, \quad \mathrm{in} \quad \mathrm{MR},
\end{aligned}
\end{equation}
where the matrices $\mathbf{H},\bar{\mathbf{H}}$ and column vectors $\mathbf{f}_0,\mathbf{f}_1,\bar{\mathbf{f}}$ are given in Appendix A.  The associated boundary conditions at $Z=0,1$ can be obtained from (\ref{eq_19}) by consistent truncations. The explicit form of boundary conditions depends on the region where the end point $Z=0$ or $Z=1$ is located. If both ends are in AR, we have 
\begin{equation}
\begin{aligned}
\label{3eq4} -\mathbf{b}^{(2)} \mathbf{U}_0|_{Z=0,1}=\gamma,\quad \mathbf{b}^{(2)} \mathbf{U}_{0Z}|_{Z=0,1}=0, \quad \mathrm{in} \quad \mathrm{AR},
\end{aligned}
\end{equation}
where the row vector $\mathbf{b}^{(2)}$ is given in Appendix A.  If the end point lies in other regions, either MR or PTR, the conditions should be modified as
\begin{equation}
\begin{aligned}
\label{3eq5}
& -\mathbf{b}^{(2)} \mathbf{U}_0=\gamma+ s_2 \xi_1 -2 s_1 \xi_2,\quad \mathbf{b}^{(2)} \mathbf{U}_{0Z}=0,\quad \mathrm{in} \quad \mathrm{MR},\\
\mathrm{or} \quad  &-\bar{\mathbf{b}}^{(2)} \mathbf{U}_0=\gamma+ (s_2 \xi_1 -2 s_1
\xi_2)\alpha_{00},\quad  \bar{\mathbf{b}}^{(2)} \mathbf{U}_{0Z}=0, \quad \mathrm{in} \quad \mathrm{PTR},
\end{aligned}
\end{equation}
where the elastic moduli $\xi_1,\xi_2$, the row vector $\bar{\mathbf{b}}^{(2)}$ and constant $\alpha_{00}$ are given in Appendix A.

Since the above simplified 1D system for $\mathbf{U}_0$ may consist of multiple regions, one needs to consider the interface between different regions and propose suitable connection conditions on interfaces, for which we turn to the next subsection.

\Remark There are many situations for the 1D system during the phase transition process. First, it may contain just one region, e.g., AR, this is the case when $\gamma$ is small with pure elastic deformations in austenite phase. Then the equation $(\ref{eq23})_1$ and condition (\ref{3eq4}) form the complete system, which leads to the unique solution of $\mathbf{U}_0$ corresponding to the homogeneous deformation. Second, for certain stress $\gamma$, the cylinder would contain multiple phase regions, e.g., AR and PTR in $(\ref{eq23})$. In this case, one needs to solve the associated system with interface conditions. Mathematically this is a free boundary problem, as the shape and position of interface are unknown and even the number of regions/interfaces is unknown beforehand.

\subsection{Connection conditions on interface}

During the phase transition process and for certain stress $\gamma$, there exist multiple phase regions in the cylinder. The interface between different regions is a free boundary. In general the shape of interface is unknown, and  the general interfaces have been investigated in \citep{songdai2015} and it turns out that the planar interface is favored (from an energy perspective). Planar interfaces are also adopted by others in 3D analytical studies \cite{zhong1, zhong2}. Here, we restrict ourselves to planar interfaces, for both simplicity and availability of analytical results, which provide more insights into the mechanism of phase transition and roles of parameters. 

\begin{figure}
\begin{center}
\includegraphics[width=3in]{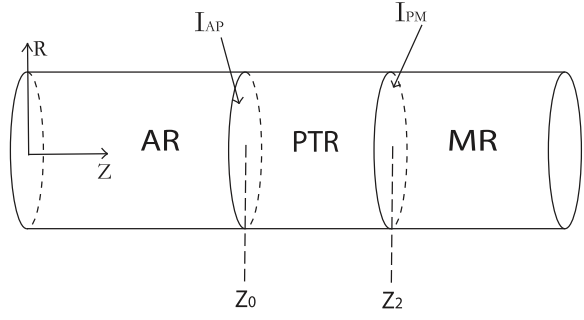}
\end{center}
\caption{Sketch of  two planar interfaces between three regions in
the reference configuration.} \label{fig1}
\end{figure}

Figure \ref{fig1} schematically illustrates two planar interfaces between three regions in the reference configuration. Interfaces $I_{AP}$ and $I_{PM}$ correspond to the surface $\alpha=0$ and $\alpha=1$, where the positions $Z_0$ and $Z_2$ of the interfaces are to be determined. For second-order vector equations in (\ref{eq23}) and considering the two unknown positions, one needs five connection conditions on each interface. These conditions can be obtained based on continuity of phase state $\alpha$ and deformation gradient $\mathbf{F}$ (see Section 5.2 of \cite{songdai2015} for details).

To present the conditions and other expressions, we first define
\begin{equation}
\begin{aligned}
\label{eq52}
& \gamma^\ast =\frac{\gamma}{E},\quad D_k^{\pm} = \frac{2B -k^{\pm} }{E},\quad \Delta \phi = \phi_2-\phi_1,\\
& D_{\phi}^+ = \frac{B+Y^{+}+ \Delta \phi}{E},\quad D_{\phi}^-= \frac{B- Y^{-}- k^-
+ \Delta \phi}{E},
\end{aligned}
\end{equation} 
where $E$ and $\nu$ are Young's modulus and Poisson's ratio. Then, $\gamma^\ast$ is the dimensionless nominal stress, and others are combinations of material parameters.

On the interface $I_{AP}$, the five connection conditions are
\begin{equation}
\begin{aligned}
\label{eq87} 
\tilde{V}_{0Z}^A (Z_0)=& \tilde{V}_{0Z}^P (Z_0),\\
W_{0Z}^{A,P}(Z_0) =& \gamma^\ast +\frac{(D_{\phi
}^{\pm}-\gamma^\ast s_2) \left(\left(2 \nu ^2+\nu +1\right) s_1-\nu
(\nu +3) s_2\right)}{2 s_1 \left(\nu   s_1-s_2\right)},\\
\tilde{V}_0^{A,P}(Z_0)=& -2 \nu \gamma^\ast-\frac{(D_{\phi
}^{\pm}-\gamma^\ast s_2) \left(4 \nu s_1+\left(\nu ^2+2 \nu -3\right)
s_2\right)}{2 s_1 \left(\nu s_1-s_2\right)},
\end{aligned}
\end{equation}
where superscripts $A$ and $P$ denote solutions in AR and PTR respectively. On the interface $I_{PM}$, we have 
\begin{equation}
\begin{aligned}
\label{eq87_1} 
\tilde{V}_{0Z}^P (Z_2)=& \tilde{V}_{0Z}^M (Z_2),\\
W_{0Z}^{P,M}(Z_2) =& \gamma^\ast +s_2+\frac{(D_{\phi
}^{\pm}-\gamma^\ast s_2 -D_k^\pm) \left(\left(2 \nu ^2+\nu +1\right) s_1-\nu
(\nu +3) s_2\right)}{2 s_1 \left(\nu   s_1-s_2\right)},\\
\tilde{V}_0^{P,M}(Z_2)=& -2 \nu \gamma^\ast -2s_1-\frac{(D_{\phi
}^{\pm}-\gamma^\ast s_2  -D_k^\pm) \left(4 \nu s_1+\left(\nu ^2+2 \nu -3\right)
s_2\right)}{2 s_1 \left(\nu s_1-s_2\right)},
\end{aligned}
\end{equation}
where superscript $M$ denotes solutions in MR.

In short, depending on the number of interfaces in the cylinder, one needs to suitably choose the connection conditions on either $I_{AP}$ or $I_{PM}$ or both, to complete the system in (\ref{eq23}-\ref{3eq5}).

\Remark In this paper, interface means the boundary between two
different regions (AR, PTR or MR), which is different from the phase interface in sharp interface models. In the present diffuse interface model, the transformation front is identified with a small region (PTR) rather than a sharp interface. This is similar to models with strain gradient terms \citep{shaw4,chang}, with a parameter for the width of a smooth PTR. In fact, the present local 3D model without strain gradient terms naturally gives a width of PTR,  determined by the material and geometric parameters. 

\section{Symmetric case and general solutions}

In the previous section, we have obtained the simplified 1D system, with possible interface conditions. Since one does not know the number and locations of regions beforehand, the governing equations and boundary conditions in the system will vary. Mathematically speaking, this is a tricky free boundary problem. The instability during phase transition is reflected by the existence of multiple solutions and the transition to a more stable solution. We recall the phase transitions process in experiments and consider a symmetric case. Then, we present the general solutions in each region, with integrating constants to be determined.

In the following we take the loading process for illustration, and unloading process can be similarly studied with slight modifications. From experiments, initially the SMA wire undergoes elastic homogeneous deformations in austenite phase. This is the case when the cylinder contains only one phase region AR, and it is easy to prove that this is a trivial case that the solution is unique, corresponding to the homogeneous deformation. At certain critical stress, the nucleation of martensite phase occurs, i.e., the transition from homogeneous deformations to inhomogeneous deformations takes place. Finally during propagation of martensite band, it is in a localized inhomogeneous deformation. Clearly, the localized inhomogeneous deformation corresponds to a case with three regions AR, PTR and MR. In between, there should be transient state of inhomogeneous deformations with two regions AR and PTR, and this state could be either energetically stable or unstable. 

Nucleation of martensite phase can occur either at end point or in the middle of cylinder, depending on the imperfection of material and experimental setup at two ends. To avoid the complicated interaction between phase interface and end surface \citep{songthesis}, we will focus on the symmetric case, where the martensite band nucleates in the middle of the cylinder. In this context, we concentrate on the interval [0, 0.5], and adopt symmetric conditions at the middle surface $Z=0.5$ in place of the end condition at $Z=1$. Since the solution is symmetric about $Z=0.5$, it is natural to adopt
\begin{equation}
\begin{aligned}
\label{3eq39} \mathbf{U}_{0Z} (0.5)=\mathbf{0}, \quad  \Leftrightarrow \quad W_{0ZZ}(0.5)=0,\quad \tilde{V}_{0Z}(0.5)=0,
\end{aligned}
\end{equation}
i.e., the two leading-order strains (the axial strain $W_{0Z}$ and the radial strain $\tilde{V}_0$) are symmetric. Actually, this is equivalent to the symmetry of profile of the cylinder, justified in the next section. As a result, we only need to examine the left half of the cylinder with (\ref{3eq39}) and the other condition (\ref{3eq4}) or (\ref{3eq5}) at $Z=0$.

In the following, we present the general solutions in each region of (\ref{eq23}). They are constructed by finding a particular solution and studying the eigenvalues of matrix $\mathbf{H}$ or $\bar{\mathbf{H}}$. The general solutions in AR ($\alpha=0$) and MR ($\alpha=1$) are given by
\begin{equation}{\fontsize{11pt}{\baselineskip}
\begin{aligned}
\label{3eq7} \mathbf{U}_{0}^A(Z)=&  \left(
\begin{array}{c} \gamma^\ast\\
 -2\nu
\gamma^\ast \end{array}\right)+ \mathbf{U}_{0c}(Z;C_1,C_2,C_3,C_4),\\
\mathbf{U}_{0}^M(Z)=&  \left(
\begin{array}{c} \gamma^\ast +
s_2\\
 -2\nu \gamma^\ast - 2s_1
 \end{array}\right) +\mathbf{U}_{0c}(Z;C_9,C_{10},C_{11},C_{12}),
\end{aligned}}
\end{equation}
where all $C_i$ ($i=1,..,4,9,..,12$) are integrating constants and
\begin{equation}{\fontsize{11pt}{\baselineskip}
\begin{aligned}
\label{3eq7_1}
 & \mathbf{U}_{0c}(Z;B_1,B_2,B_3,B_4) = B_1 e^{-d_1
\tilde{Z}}\left(
\begin{array}{c} \cos(d_2 \tilde{Z}) \\
q_1 \cos(d_2 \tilde{Z}) + q_2 \sin(d_2 \tilde{Z})\\
\end{array}\right)\\
&+ B_2 e^{-d_1 \tilde{Z}}\left(
\begin{array}{c} -\sin(d_2 \tilde{Z}) \\
-q_1 \sin(d_2 \tilde{Z}) + q_2 \cos(d_2 \tilde{Z})\\
\end{array}  \right)\\
&+ B_3 e^{d_1 \tilde{Z}}\left(\begin{array}{c}
\cos(d_2 \tilde{Z}) \\
q_1 \cos(d_2 \tilde{Z}) - q_2 \sin(d_2 \tilde{Z})\\
\end{array}\right)\\
&+ B_4 e^{d_1 \tilde{Z}}\left(\begin{array}{c}
\sin(d_2 \tilde{Z}) \\ q_1 \sin(d_2 \tilde{Z}) + q_2 \cos(d_2 \tilde{Z})\\
\end{array} \right),
\end{aligned}}
\end{equation}
where $\tilde{Z}= Z-Z_0$, and the explicit expressions of $d_1,d_2,q_1,q_2$ are given in Appendix B. In the solution, we have made a shift in $Z$ by $Z_0$ (any shift will be a solution), which makes it easier to determine the integrating constants in later sections. The general solution in PTR is
\begin{equation}
\begin{aligned}
\label{eq59} \mathbf{U}^P_{0}(Z)=\mathbf{U}^{P}_{0p} + C_5
\bar{\mathbf{q}}_1 e^{d_3 \tilde{Z}} + C_6 \bar{\mathbf{q}}_1 e^{-d_3 \tilde{Z}} + C_7 \bar{\mathbf{q}}_2 \cos( d_4 \tilde{Z}) + C_8 \bar{\mathbf{q}}_2 \sin( d_4 \tilde{Z}),
\end{aligned}
\end{equation}
where $C_i$ ($i=5,..,8$) are integrating constants and
\begin{equation}
\begin{aligned}
\label{eq60}
& \mathbf{U}_{0p}^{P}=\left(\frac{D_{\phi }^{\pm}
s_2}{D_k^{\pm}}+\frac{\gamma^\ast \left(D_k^{\pm}
-s_2^2\right)}{D_k^{\pm}},\quad -\frac{2 D_{\phi}^{\pm}
s_1}{D_k^{\pm}}-\frac{2 \gamma^\ast  \left(\nu  D_k^{\pm} -s_1
s_2\right)}{D_k^{\pm} }\right)^T, \\
& \bar{\mathbf{q}}_1=[1,\ \bar{q}_{1}]^T, \qquad
\bar{\mathbf{q}}_2=[1,\ \bar{q}_{2}]^T.
\end{aligned}
\end{equation}
The constants $d_3,d_4, \bar{{q}}_1,\bar{{q}}_2$ are given in Appendix B.

\begin{figure}
\begin{center}
\includegraphics[width=3in]{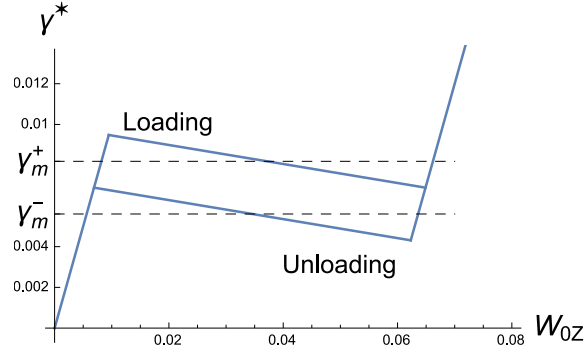}
\end{center} \caption{1D homogeneous stress-strain curve
and Maxwell stresses $\gamma_m^\pm$ for the loading and unloading processes.} \label{fig2}
\end{figure}

Based on the particular solutions ($C_i =0$) in (\ref{3eq7}, \ref{eq59}, \ref{eq60}), Figure \ref{fig2} shows the 1D stress-strain (axial strain) relation for homogeneous states of the loading and unloading processes, where the Maxwell stresses $\gamma_m^{\pm}$ are defined by
\begin{equation}
\begin{aligned}
\label{eq51_1} \gamma_m^{\pm} =\frac{
2D_{\phi}^{\pm}-D_{k}^{\pm}}{2s_2}.
\end{aligned}
\end{equation}
Figure \ref{fig2} shows the strain-softening behaviour, with reasonable material constants in Appendix C. Each point in Figure  \ref{fig2} represent a solution for homogeneous deformation. These are the unique solutions when there is only one region, AR, PTR or MR at different stress $\gamma^\ast$. However, the homogeneous solution with pure PTR region, corresponding to the softening branch in the curve,  is locally unstable from an energy point of view \citep{Ericksen}. This part of stress-strain curve will not be observed in experiments. In the next sections, we will focus on inhomogeneous deformations where multiple regions are present, and investigate the phase transition process. For inhomogeneous deformations, we restrict ourselves to the possibly smallest number of interfaces and regions, as adding more regions and interfaces will make solutions unfavored energetically. Since the general solution is given in each region, determining the solutions for inhomogeneous deformations is equivalent to determining the associated integrating constants.

\section{Two-region solutions}

\begin{figure}
\centerline{\includegraphics[width=3.5in]{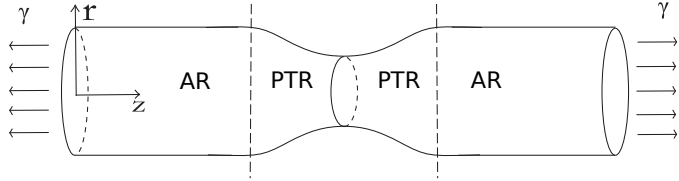}} \caption{
Current configuration with two regions AR and PTR in the symmetric
case.}\label{3fig8}
\end{figure}

In this section, we study the case when the cylinder contains only two regions in $[0,0.5]$, AR and PTR from left to right (see Figure \ref{3fig8}). We first present the analytical results for the two-region solutions for given stress $\gamma^\ast$, and then calculate the total elongations for the solutions, which are often measured in displacement-controlled experiments on SMAs .

\subsection{Analytical results for given $\gamma^\ast$}

In this case, the general solution $\mathbf{U}_0$ in $[0,0.5]$ is composed of  $(\ref{3eq7})_1$ and (\ref{eq59}), which contains eight integrating constants $C_1,..,C_8$ and one unknown position $Z_0$ from the interface $I_{AP}$. The two boundary conditions in (\ref{3eq4}) at $Z=0$, two symmetric conditions in (\ref{3eq39}) at $Z=0.5$ and five connection conditions $(\ref{eq87})$ at interface $Z=Z_0$ can totally determine the nine unknowns.

First, the integrating constants $C_1,..,C_8$ can be expressed in terms of $Z_0,\gamma^\ast$ and material parameters. Since the transition region is in the middle of the cylinder, the terms containing $C_3$ and $C_4$ (with factor $e^{-d_1 Z_0}$) in solutions $(\ref{3eq7})_1$ at the left end are exponentially small.
As a result, the two boundary conditions (\ref{3eq4}) at $Z=0$ lead to $C_1= C_2
=0$. Then, substituting $(\ref{3eq7})_1$ into conditions (\ref{eq87}) leads to
\begin{equation}
\begin{aligned}
\label{3eq11} C_3 = \left(D_{\phi }^{+} -\gamma^\ast s_2\right)
q_{11},\quad C_4 = \left(D_{\phi }^{+} - \gamma^\ast s_2\right)q_{22},
\end{aligned}
\end{equation}
and the conditions in $(\ref{eq87})_{2,3}$ and (\ref{3eq39}) imply
\begin{equation}
\begin{aligned}
\label{3eq44} C_5=&\frac{ -\bar{q}_{22} \left(D_{\phi }^+ -\gamma^\ast
s_2\right) } {\left(1+e^{2  d_3   \bar{Z}_0 }\right)}, \quad
C_6=\frac{-\bar{q}_{22} e^{2 d_3 \bar{Z}_0 } \left(D_{\phi }^+
-\gamma^\ast  s_2\right) }{\left(1+e^{2
d_3   \bar{Z}_0 }\right)},\\
C_7=& \bar{q}_{11} \left(D_{\phi }^+ -\gamma^\ast  s_2\right),\quad
C_8= \bar{q}_{11} \tan ( d_4 \bar{Z}_0 ) \left(D_{\phi }^+
-\gamma^\ast s_2\right),
\end{aligned}
\end{equation}
where $\bar{Z}_0 = 0.5-Z_0$ is the width of PTR in this case, and the material constants $q_{11}, q_{22}, \bar{q}_{11},\bar{q}_{22}$ are defined in Appendix B.

\Remark From general solutions of $\mathbf{U}_0$ together with the above results and recursive relations, we have the following equivalence
\begin{equation}
\begin{aligned}
\label{3eq43}
&W_{0ZZ}(0.5)=0,\quad \tilde{V}_{0Z}(0.5)=0, \\
\Leftrightarrow \quad & \tilde{V}_{Z}(0.5,R)\approx \frac{1}{2} \tilde{V}_{0Z}(0.5) + \frac{1}{4}R^2 \tilde{V}_{1Z}(0.5) =0.
\end{aligned}
\end{equation}
Thus, the symmetric conditions of the two strains $W_{0Z}$ and $\tilde{V}_{0}$ are equivalent to the symmetric conditions of the profile (radial displacement) at all points on the middle surface. This equivalence also holds for the three-region solutions in the next section, so there is no need to distinguish the two sets of symmetric conditions.

Next, by substituting the above expressions of $C_1,..,C_8$ into the last condition $(\ref{eq87})_{1}$,  we obtain an equation for $\bar{Z}_0$ (note $\bar{Z}_0= 0.5-Z_0$) in the form
\begin{equation}
\begin{aligned}
\label{3eq46} \left(D_{\phi }^+ -\gamma^\ast  s_2\right) \left\{-n_{11} \tan [d_4 \bar{Z}_0] + n_{12} \tanh [d_3 \bar{Z}_0] + n_{13} \right\} =0,
\end{aligned}
\end{equation}
where the material constants $n_{11},n_{12}, n_{13}$ are independent of $\gamma^\ast$ and are given in Appendix B. Thus for $\gamma^\ast< \gamma_{NM}^\ast : = D_{\phi }^+ /s_2$,  the unknown $\bar{Z}_0$ or interface position $Z_0$ is determined by the nonlinear equation in curly brackets in (\ref{3eq46}) and does not depend on the given parameter $\gamma^\ast$. Therefore, determining the two-region solutions has reduced to solving a nonlinear equation for $\bar{Z}_0$.

In order to get closed-form solutions of $\bar{Z}_0$ (and hence $\mathbf{U}_0$), we neglect the exponentially small terms of $O(e^{-d_3 \bar{Z}_0})$ in (\ref{3eq46}) and obtain
\begin{equation}
\begin{aligned}
\label{3eq47} \bar{Z}_0^{(k)} = \frac{1}{d_4}\left(\arctan \left(
\frac{n_{12} + n_{13}}{n_{11}} \right) +k \pi \right), \quad k=1,2...,
\end{aligned}
\end{equation}
where $\bar{Z}_0^{(k)}$ is the $k^{th}$ solution for $\bar{Z}_0$, and
$-\pi/2 < \arctan x < \pi/2$. This formula is numerically verified by solving the whole system of $C_1,..,C_8,Z_0$, with chosen material constants in (\ref{eq75_1}) and $a=0.03$. The first few solutions are easily obtained by Newton's method, and the solution with smallest width $\bar{Z}_0$ is given in Appendix C.  By comparison with the direct numerical results, the above analytical approximation (\ref{3eq47}) is very accurate, e.g.,
\begin{equation}
\begin{aligned}
\label{3eq48}
&\mathrm{numerical}: \quad &\bar{Z}_0^{(1)}=0.0570056, \quad \bar{Z}_0^{(2)} =0.111648,\\
&\mathrm{analytical}: \quad &\bar{Z}_0^{(1)} =0.0570109, \quad \bar{Z}_0^{(2)}
=0.111648.
\end{aligned}
\end{equation}
The relative error of $\bar{Z}_0^{(k)}$ is almost 0 for any $k\geq 1$. For a given $\gamma^\ast$ in certain interval, each $\bar{Z}_0^{(k)}$ determines one solution of $\mathbf{U}_0$, called the period-k solution.

Finally, based on the result (\ref{3eq47}), we analyze the dependence of the width of PTR on material and geometrical parameters. By using the small parameter $\epsilon_1:= D_k^{+}/s_2^2 $ and retaining two leading terms, we obtain
\begin{equation}
\begin{aligned}
\label{3eq50} \frac{\bar{Z}_0^{(k)}}{a}=& \frac{(4 k+1) \pi  s_1}{8
\sqrt[4]{\epsilon_1} \sqrt{s_2} \sqrt[4]{s_1^2-2 \nu  s_2 s_1+s_2^2}} +
\frac{ \sqrt[4]{\epsilon_1 } (\nu +1) \sqrt{s_2} }{32 \left(\nu
s_1-s_2\right)
\left(s_1^2-2 \nu s_2 s_1+s_2^2\right)^{3/4}} \\
&[((-4 \pi  k - \pi -12) \nu +8) s_1^2+((4 k + 1) \pi (\nu +1)-12 (\nu
-1)) s_2 s_1 \\ &+(-4 \pi  k - \pi +4) s_2^2].
\end{aligned}
\end{equation}
Clearly, the width $\bar{Z}_0^{(k)}$ for any $k$ is proportional to radius $a$, and increases with $k$. The proportional dependence on geometric parameter $a$ is expected in both experiments and theoretical works \citep{chang,he1}. With $k=1$ and the condition $s_1=s_2/2$ (volume preserving condition for the natural configuration $\mathbf{G}$ in (\ref{eq3})), it reduces to
\begin{equation}
\begin{aligned}
\label{3eq51} \frac{\bar{Z}_0^{(1)}}{a}=&\frac{5 \pi }{8
\sqrt{2} \sqrt[4]{5-4 \nu } \sqrt[4]{\epsilon_1 }}+\frac{ (5\pi  (\nu -2)-
36 \nu +48) (\nu +1) \sqrt[4]{\epsilon_1} }{16 \sqrt{2} (5-4 \nu )^{3/4}
(\nu -2) }.
\end{aligned}
\end{equation}
It depends on material parameters mainly through $\epsilon_1$ with leading power $-1/4$, whereas the Poisson's ratio $\nu$ only has minor effect. Thus to leading order, the width $\bar{Z}_0^{(1)}$ is proportional to  the square root of axial transformation strain $\sqrt{s_2}$, and it becomes larger as $D_k^{+}$ decreases. From the experiments reported in \cite{sun5}, the width of PTR (transformation front) is closely related to the stress drop of the stress-strain curve, and tends to increase as the stress drop decreases. Here the term $D_k^{+}$ defined in (\ref{eq52}) is associated with stress drop or strain softening behavior, and is determined by the competition between the interfacial energy (constant $B$) and the dissipation (constant $k^+$). It is shown in \citep{songdai} that stress drop decreases with decrease of $D_k^{+}$.  Therefore, the qualitative feature of the analytical formula agrees well with experiments. In addition, the appearance of this natural width is due to both the strain-softening behavior and the geometric effect, since the high-dimensional effect is retained in the formulation (hidden in recursive formulas).

\subsection{Period-1 solution and the total elongation}

In the previous subsection, period-k solutions for given $\gamma^\ast$ are obtained. Actually, for a displacement-controlled process as in experiments, the parameter $\gamma^\ast$ is to be determined by a given total elongation based on energy criterion. One can expect that the solution with larger width $\bar{Z}_0$ would have larger total energy (defined in (\ref{eq76}), see also \cite{songdai2015}). This subsection then focuses on the period-1 solution with smallest width $\bar{Z}_0^{(1)}$, and presents the total elongation of such solution with varying $\gamma^\ast$.

First, we present the solutions of the leading-order axial strain $W_{0Z}$ and phase state $\alpha$ for period-1 solution and some given $\gamma^\ast$. They can be represented as
\begin{equation}
\begin{aligned}
\label{3eq52} &W_{0Z}(Z;\gamma^\ast) = \hat{f}_w(Z) (
\gamma^\ast_{NM} - \gamma^\ast )
+\gamma^\ast,\\
&\alpha(R,Z;\gamma^\ast) =  \hat{f}_\alpha(R,Z) (
\gamma^\ast_{NM} - \gamma^\ast ),
\end{aligned}
\end{equation}
where $\hat{f}_w$ and $\hat{f}_\alpha$ are some shape functions independent of
$\gamma^\ast$ and $D_\phi^{+}$ (recall $\gamma^\ast_{NM}=D_\phi^{+}/s_2$), whose expressions are omitted for brevity. The period-1 two-region solutions are obtained only for $\gamma^\ast$ in certain interval, since the phase state is restricted by $0<\alpha <1$, which leads to the range
\begin{equation}
\begin{aligned}
\label{3eq53} \gamma_{c}^\ast < \gamma^\ast < \gamma^\ast_{NM}.
\end{aligned}
\end{equation}
For the chosen material constants in (\ref{eq75_1}) we have
\begin{equation}
\begin{aligned}
\label{3eq53_1}
\gamma_{c}^\ast=0.00801694 < \gamma_m^+ = 0.00818966, \quad \gamma^\ast_{NM} = 0.00948276.
\end{aligned}
\end{equation}

\begin{figure}
\begin{center}
\subfigure[]{\includegraphics[width=2.5in]{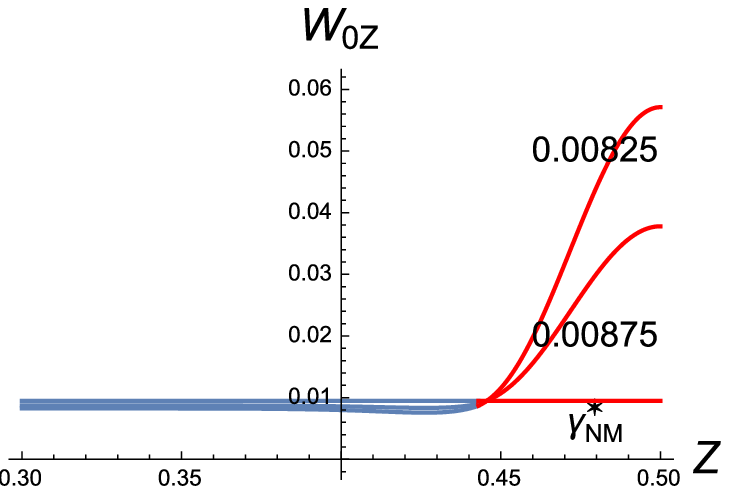} }
\subfigure[]{\includegraphics[width=2.5in]{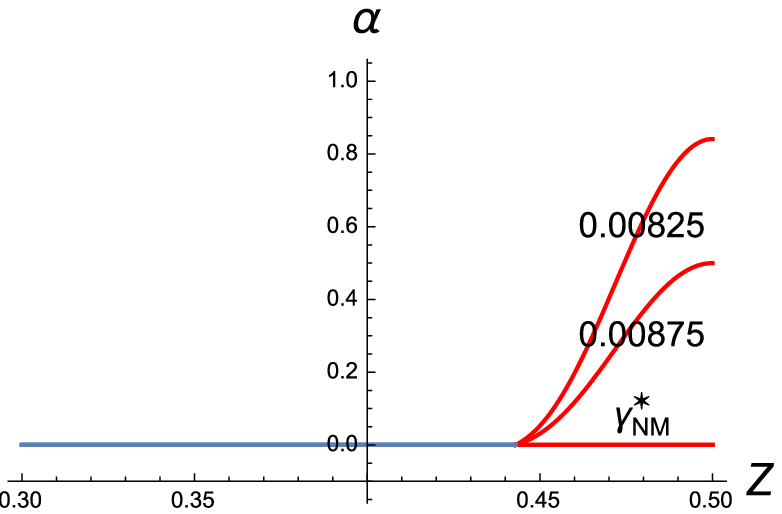}}
\end{center}
\caption{The solution profiles with varying $\gamma^\ast$ shown near curves for (a) leading order axial strain $W_{0Z}$, (b) phase state $\alpha(0,Z)$.} \label{3fig9}
\end{figure}

Figure \ref{3fig9} shows the profiles of $W_{0Z}(Z)$ and $\alpha(0,Z)$ in the interval $[0.3,0.5]$ for varying $\gamma^\ast$, indicated near the curves. The red curves represent the solutions in PTR, and blue curves represent solutions in AR and extend as a constant until left end $Z=0$.  It is shown that the width of PTR is fixed independent of $\gamma^\ast$, while the maximum values of the curves at $Z=0.5$ increases with decease of $\gamma^\ast$. As $\gamma^\ast$ varies, the shape of the curves for $W_{0Z}$ or $\alpha$ in PTR remains the same, characterized by the function $\hat{f}_w$ or $\hat{f}_\alpha$. 

The fixed width  $\bar{Z}_0^{(1)}$ of PTR is an important feature of two-region solution, and its dependence on parameters has been discussed in the previous subsection. As we will see in Section \ref{sec7} that the solution will transition from a homogeneous deformation (solution) to a period-1 two-region solution during phase transition. This implies that phase transition always takes place simultaneously in a fixed PTR region of width $\bar{Z}_0^{(1)}$ rather than at a point. Importantly, this is obtained by naturally keeping high-dimensional effect in the 3D local model instead of introducing strain gradient terms \cite{chang}.

\begin{figure}
\begin{center}
\includegraphics[width=2.5in]{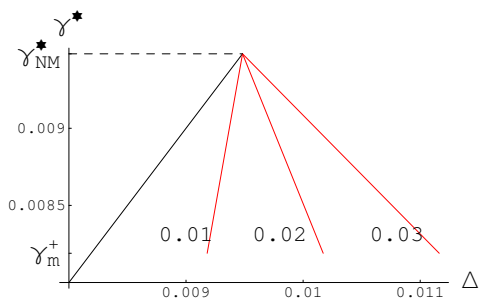}
\end{center}
\caption{ The stress-elongation curves of the two-region solutions (in red lines)
with different radius $a=0.01,0.02,0.03$.}
\label{3fig10}
\end{figure}

The total elongation is defined by
\begin{equation}
\begin{aligned}
\label{3eq54} \Delta = \int_0^{1} W_{0Z} (Z) \mathrm{dZ}= 2
\int_0^{Z_0} W_{0Z}^A (Z) \mathrm{dZ} + 2 \int_{Z_0}^{1/2} W_{0Z}^P
(Z) \mathrm{dZ}.
\end{aligned}
\end{equation}
By (\ref{3eq52})  it can be expressed in the form
\begin{equation}
\begin{aligned}
\label{3eq55} \Delta =  K_1 (\gamma^\ast - \gamma^\ast_{NM})
+\gamma^\ast_{NM}, \quad K_1 = 1 -  \int_0^1  \hat{f}_w(Z) dZ =
1 - K_2 a,
\end{aligned}
\end{equation}
where $K_2$ only depends on material constants (not on $a$). The explicit expression of $K_2$ is omitted for brevity, but its approximation is given below. For the chosen material values in (\ref{eq75_1}), we have $K_2= 76.6342$. Figure \ref{3fig10} shows the stress-elongation curves of two-region solutions for three different radii $a$ in red lines, where the reference black line corresponds to homogeneous deformation in pure austenite phase (cf. Figure \ref{fig2}).  Clearly, each curve is a line, passing through the peak point $(\gamma^\ast_{NM},\gamma^\ast_{NM})$. As the radius $a$ tends to 0, the red line of the two-region solution tends to the black line $\gamma^\ast = \Delta$. As the radius $a$ increases, the line will rotate to the right with the peak point fixed. There is a critical radius $a_c=1/K_2$ (between $0.01$ and $0.02$ here) such that the red line is vertical, which is crucial for the stability later.

To get an idea of the roles of material constants on elongation, we expand $K_2$ in terms of the small parameter $\epsilon_1$ defined above (\ref{3eq50}).  With $s_1=s_2/2 $, we get the two-term approximation
\begin{equation}{\fontsize{10pt}{\baselineskip}
\begin{aligned}
\label{3eq56} K_2 =&\frac{\left(5 \pi +e^{5 \pi /2} (-4+5 \pi
)\right)}{4 \sqrt{2} \left(1+e^{5 \pi /2}\right)
\sqrt[4]{5-4 \nu } \epsilon_1^{5/4}} + \frac{ 1}{8 \sqrt{2}
\left(1+e^{5 \pi /2}\right)^2
(5-4 \nu )^{3/4} (\nu -2) \epsilon_1^{3/4}}\\
&[4 \left(\nu ^2-25 \nu +28\right)+5 e^{5 \pi } \pi  \left(\nu ^2-\nu
-2\right)+5 \pi  \left(\nu ^2-\nu -2\right)\\
&-2 e^{5 \pi /2} \left(-38 \nu ^2+62 \nu +5 \pi  \left(\nu ^2-\nu
-2\right)-8\right)],
\end{aligned}}
\end{equation}
which means $K_2$ depends on material parameters mainly through $\epsilon_1$ with leading power $-5/4$, while Poisson's ratio $\nu$ has little impact. This implies $a_c \sim \epsilon_1^{5/4}$ to leading order. In particular, as $D_k^+$ decreases, $K_2$ will increase and thereby $K_1$ will decrease by (\ref{3eq55}) for fixed $a$, thus the line of two-region solution in Figure \ref{3fig10} will rotate to the right with fixed peak point. 

\Remark As mentioned in the previous subsection, $D_k^{+}$ is crucial parameter associated with stress drop or strain softening behavior. As $D_k^+$ tends to 0, $K_2$ tends to $\infty$ and the red line of the two-region solution in Figure \ref{3fig10} tends to a horizontal line $\gamma^\ast= \gamma^\ast_{NM}$, together with an infinite PTR. This agrees very well with experiments in \cite{sun5}. This makes sense since if there is no strain softening behaviour, there is no need to study the two-region solutions and the horizontal line is just the critical case of the stable homogeneous state with only PTR (the second branch of stress-strain curve in Figure \ref{fig2} is no longer decreasing).

\section{Three-region solutions}

\begin{figure}
\centerline{\includegraphics[width=3.5in]{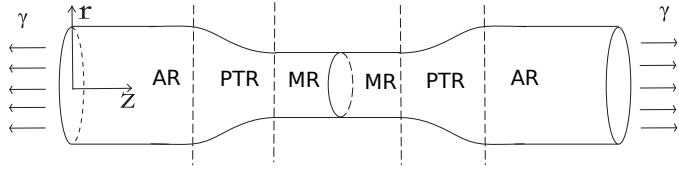}} \caption{
Current configuration of the cylinder with three regions in symmetric case.}\label{3fig11}
\end{figure}

In this section, we investigate the case with three regions, i.e., AR, PTR and MR  in the left half of the cylinder (see Figure \ref{3fig11}). This corresponds to the localized inhomogeneous deformations during propagation of phase transformation front in experiments. The previous work in \cite{songdai2015} is a special (ideal) case when the MR region is relatively large or PTR is far away from middle surface. The results here are much richer than the previous one, and the solutions are quite sensitive with small MR. We first present the semi-analytical results for the solution with given $\gamma^\ast$, and then compute the total  elongation.

In this case, the general solution $\mathbf{U}_0$ in $[0,0.5]$ is composed of  (\ref{3eq7}) and (\ref{eq59}), which contains twelve integrating constants $C_1,..,C_{12}$ and two unknown positions $Z_0$ and $Z_2$ for interfaces. The boundary conditions (\ref{3eq4}) at $Z=0$, symmetric conditions (\ref{3eq39}) at $Z=0.5$ and ten connection conditions (\ref{eq87}) and (\ref{eq87_1}) at interfaces can totally determine the unknowns.

The integrating constants $C_1,..,C_{12}$ can be expressed by $\gamma^\ast,Z_0,Z_2$ and material constants. The constants $C_1,..,C_4$ are the same as the previous section (see (\ref{3eq11})). By the expressions $(\ref{eq87})_{2,3}$ and $(\ref{eq87_1})_{2,3}$, the constants $C_5,..,C_8$ are given by
\begin{equation}
\begin{aligned}
\label{3eq30} C_5 = &  \frac{\bar{q}_{22} \left(e^{d_3  \bar{Z}_2 }
D_k^+-\left(-1+e^{d_3 \bar{Z}_2 }\right) \left(D_{\phi
}^{+}-\gamma^\ast s_2\right)\right) }{ \left(-1+e^{2 d_3 \bar{Z}_2
}\right) },\\
C_6 = &\frac{\bar{q}_{22} e^{d_3  \bar{Z}_2 }
\left(-D_k^+-\left(-1+e^{d_3 \bar{Z}_2 }\right) \left(D_{\phi
}^{+}-\gamma^\ast s_2\right)\right) }{\left(-1+e^{2 d_3 \bar{Z}_2
}\right) },\\
C_7 = &\bar{q}_{11}\left(D_{\phi }^{+}-\gamma^\ast s_2\right), \\
C_8= &-\bar{q}_{11} \csc (d_4  \bar{Z}_2 ) \left(D_k^++(\cos
(d_4 \bar{Z}_2 )-1) \left(D_{\phi }^{+}-\gamma^\ast s_2\right)\right),\\
\end{aligned}
\end{equation}
where $\bar{Z}_2 = Z_2 - Z_0$ is the width of PTR in this case, and the material constants $\bar{q}_{11},\bar{q}_{22}$ are given in Appendix B. One can similarly obtain the expressions of $C_9,..,C_{12}$, given in Appendix B.

Then for a given $\gamma^\ast$,  the system with fourteen unknowns has reduced to a system of two unknowns $\bar{Z}_2$ and $\bar{Z}_0$ (recall $\bar{Z}_0= 0.5-Z_0$), determined by the two conditions $(\ref{eq87})_{1}$ and $(\ref{eq87_1})_{1}$. First from $(\ref{eq87})_{1}$, the width $\bar{Z}_2$ of PTR is determined by
\begin{equation}
\begin{aligned}
\label{3eq34} f_2 (\bar{Z}_2;\gamma^\ast) = & \hat{m}_1 (\bar{Z}_2) (\gamma^\ast-\gamma_m^+) + \hat{n}_1(\bar{Z}_2) =0,
\end{aligned}
\end{equation}
where
\begin{equation}
\begin{aligned}
\label{3eq34_1}
& \hat{m}_1(Z_2) =\frac{s_2}{D_{k}^{\pm}}
\left(n_{11} \tan\left(\frac{1}{2}d_4 Z_2\right)- n_{12}
\tanh\left(\frac{1}{2}d_3 Z_2\right) - n_{13}\right),\\
& \hat{n}_1(\bar{Z}_2) = \frac{1}{2} \left( n_{11} \cot\left(\frac{1}{2}d_4 \bar{Z}_2\right)+  n_{12} \coth\left(\frac{1}{2}d_3 \bar{Z}_2\right) +  n_{13} \right),
\end{aligned}
\end{equation}
and $n_{11},n_{12},n_{13}$ are  given in Appendix B. Numerical evidence shows that there exist solutions only for $\gamma^\ast$ in a small neighbourhood of $\gamma_m^+$.  For the special case $\gamma^\ast = \gamma_m^+$, the equation (\ref{3eq34}) reduces to $\hat{n}_1(\bar{Z}_2) =0$, which is exactly the infinite case in \citep{songdai2015}. For this special case, we recall that the approximate period-k solution is given by
\begin{equation}
\begin{aligned}
\label{eq99_0} \bar{Z}_2^{(k)} = \frac{2}{d_4}\left( \mathrm{arccot}\left(-\frac{n_{12}+n_{13}}{n_{11}}  \right) +k \pi\right),\quad k=1,2...,
\end{aligned}
\end{equation}
where $-\pi<\mathrm{arccot}\,  x < 0$. It is verified by direct numerical results \cite{songdai2015} that this formula is very accurate, with less than 1\% error for all $k$. 

\begin{figure}
\begin{center}
\subfigure[$\gamma^\ast=0.00819$]{\includegraphics[width=2.5in]{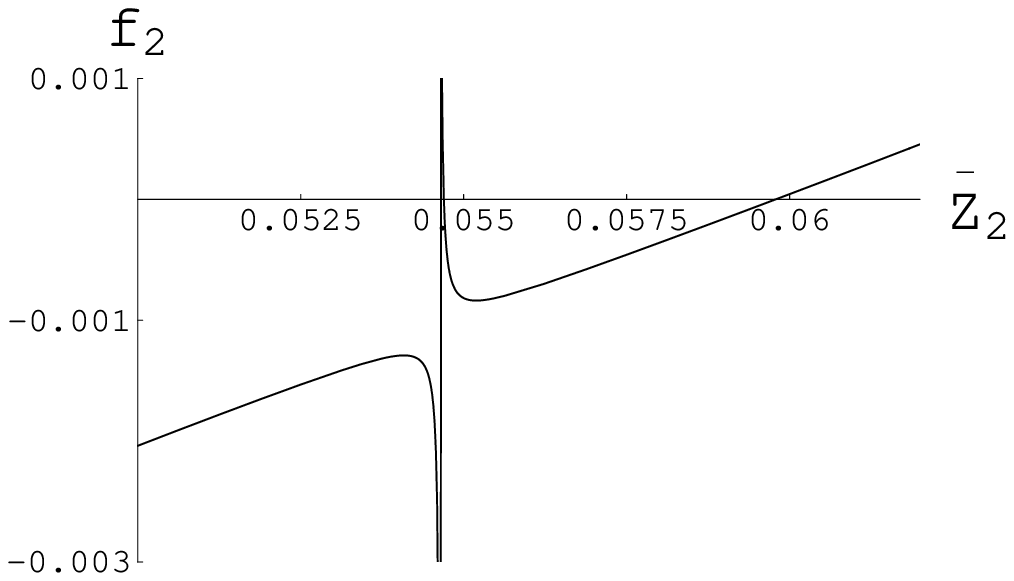}\label{3fig5a}}
\subfigure[$\gamma^\ast=\gamma_m^+= 0.00818966$]{\includegraphics[width=2.5in]{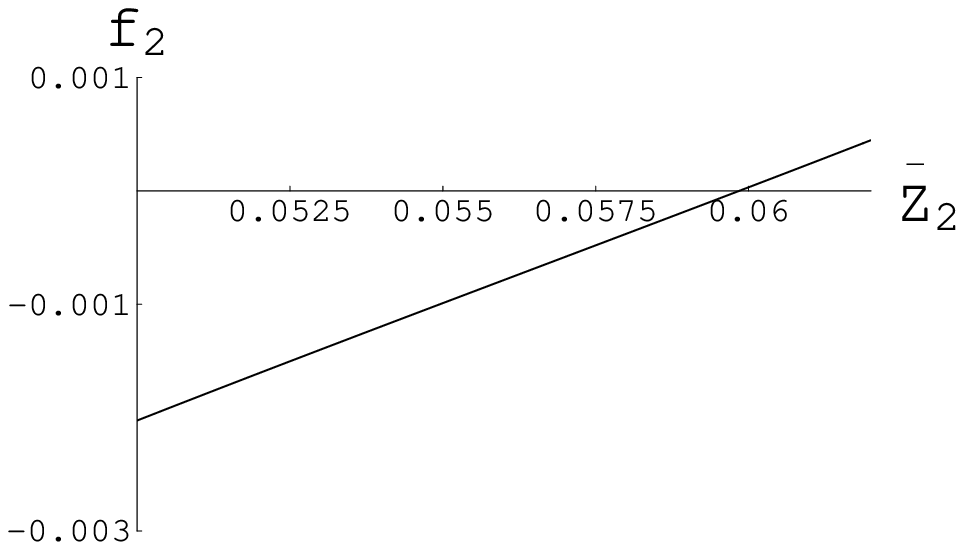}\label{3fig5b}}
\end{center} \caption{ The function $f_2(\bar{Z}_2,\gamma^\ast)$ with two different $\gamma^\ast$, $a=0.03$ and the material constants in (\ref{eq75_1}).}\label{3fig5}
\end{figure}

For the general case $\gamma^\ast \ne \gamma_m^+$, there are singularities in $f_2 (\bar{Z}_2;\gamma^\ast)$ due to the term $\tan(d_4 \bar{Z}_2/2)$ in $\hat{m}_1$. There are two situations. First, one set of period-k solutions is just a small perturbation of (\ref{eq99_0}), since $\gamma^\ast$ is close to $\gamma_m^+$. One can easily derive the regular period-k solutions 
\begin{equation}
\begin{aligned}
\label{eq99} \bar{Z}_{2,r}^{(k)} & = \frac{2}{d_4}\left( \mathrm{arccot}\left(-\frac{n_{12}+n_{13,r}}{n_{11}}  \right) +k \pi\right),\quad k=1,2...,\\
n_{13,r} & = n_{13} + 2 \hat{m}_1 (\bar{Z}_{2}^{(k)}) (\gamma^\ast-\gamma_m^+).
\end{aligned}
\end{equation}
Second, there are extra period-k solutions for $\bar{Z}_2$ near the singularities, namely,
\begin{equation}
\begin{aligned}
\label{eq99_1} \bar{Z}_{2,s}^{(k)} \approx \frac{(2k-1)}{d_4} ,\quad k=1,2, ...
\end{aligned}
\end{equation}
Figure \ref{3fig5a} shows the curve of $f_2$ with $\gamma^\ast=0.00819$, $a=0.03$ and the material constants in  (\ref{eq75_1}), indicating that the first singularity is near $0.055$. The two roots near $0.055$ and $0.06$ correspond to the two period-1 solutions of $\bar{Z}_2$ with $k=1$ in (\ref{eq99}) and (\ref{eq99_1}). Figure \ref{3fig5b} shows that the singularity disappears for the special case $\gamma^\ast= \gamma_m^+$.

Next by $(\ref{eq87_1})_{1}$, the equation for determining $\bar{Z}_0$ can be written as
\begin{equation}
\begin{aligned}
\label{3eq61} f_1(\bar{Z}_0,\bar{Z}_2, \gamma^\ast)= \frac{s_2}{D_k^+} \hat{m}_3(t) \left(\gamma^\ast -
\frac{D_{\phi }^{+}- D_k^+}{s_2}\right) + 2 \hat{m}_1 (\bar{Z}_2)
\hat{n}_3(t) (\gamma^\ast - \gamma_m^+)=0,
\end{aligned}
\end{equation}
where $t:=\bar{Z}_0 - \bar{Z}_2$ is the width of MR, and the functions $\hat{m}_3(t)$ and $\hat{n}_3(t)$
are given by
\begin{equation}
\begin{aligned}
\label{3eq62}
&\hat{m}_3(t)= n_{13} e^{-2 d_1 t } + n_{14} \sin(2 d_2
t) + n_{13} \cos(2 d_2 t),\\
&\hat{n}_3(t)= \cos(2 d_2 t) + \cosh(2 d_1 t).
\end{aligned}
\end{equation}
It is not straightforward to obtain analytical solutions of $\bar{Z}_0$ from (\ref{3eq61}), instead we discuss the typical features of this equation for given $\gamma^\ast$ and given period-1 solution of $\bar{Z}_2$ (either $\bar{Z}_{2,r}^{(1)}$ in (\ref{eq99}) or $\bar{Z}_{2,s}^{(1)}$ in (\ref{eq99_1})). Corresponding to $\bar{Z}_{2,r}^{(1)}$ in (\ref{eq99}), one can obtain two solutions for $\bar{Z}_0$. The two sets of solutions for $(\bar{Z}_0,\bar{Z}_2)$ are verified by the numerical results of solving the whole nonlinear system of 14 equations with Newton's method. No solution for $\bar{Z}_0$ is found corresponding to $\bar{Z}_{2,s}^{(1)}$ in (\ref{eq99_1}). In addition, when $\gamma^\ast$ is extremely close to $\gamma_m^+$ (say $<10^{-8}$), more than two solutions of $\bar{Z}_0$ are found corresponding to $\bar{Z}_{2,r}^{(1)}$ in (\ref{eq99}). In this case, the solutions are quite sensitive to $\gamma^\ast$ due to the term $(\gamma^\ast -\gamma_m^+)\hat{n}_3(t)$ in $(\ref{3eq61})$. This is certainly expected, since for infinitely long cylinder in \cite{songdai2015} we always get  $\gamma^\ast= \gamma_m^+$, and thus any relatively large $\bar{Z}_0$ with $\gamma^\ast$ extremely close to $\gamma_m^+$ is always a solution with exponentially small error.

\begin{figure}
\begin{center}
\subfigure[]{\includegraphics[width=2.5in]{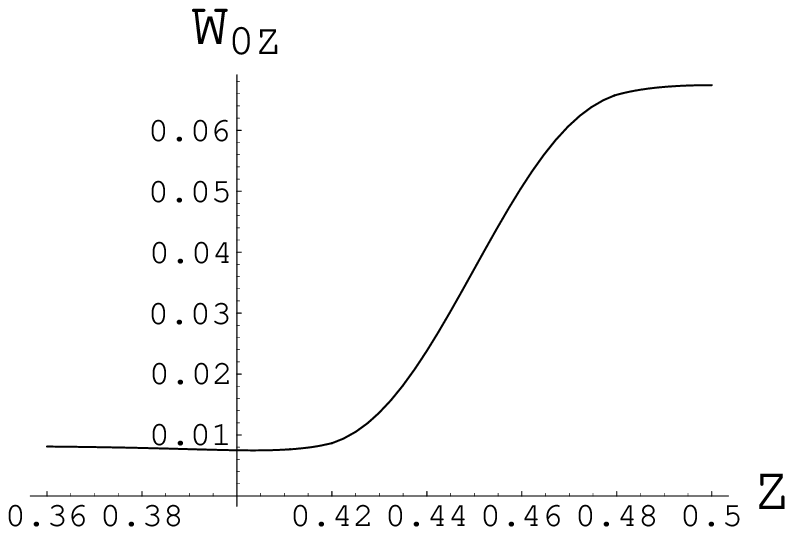}\label{3fig12a}}
\subfigure[]{\includegraphics[width=2.5in]{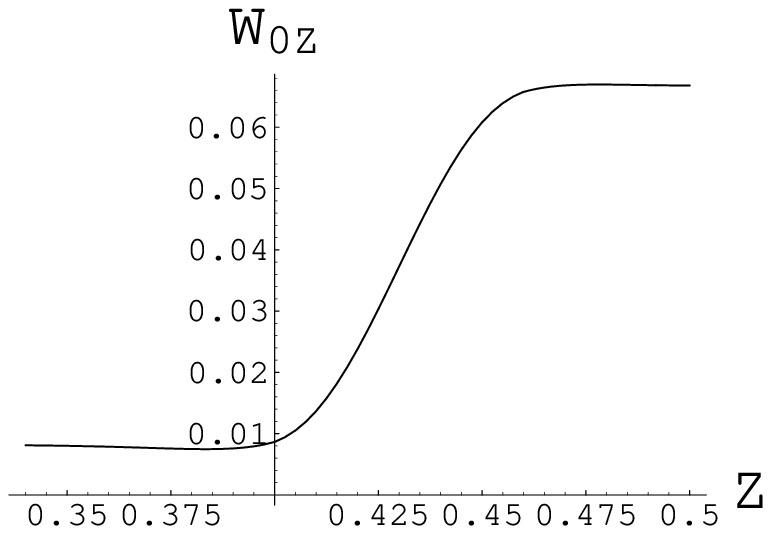}\label{3fig12b}}
\end{center}
\caption{ The leading order axial strain (a) with $\bar{Z}_0=0.08$, $\gamma^\ast=0.00819143$ and $\bar{Z}_2 =
0.0595209$, (b) with $\bar{Z}_0=0.1$, $\gamma^\ast=0.00818997$ and
$\bar{Z}_2 = 0.0597928$.}\label{3fig12}
\end{figure}

To sum up, for a given $\gamma^\ast$, the solution sets of $(\bar{Z}_2,\bar{Z}_0)$ can be determined by a system of two algebraic equations (\ref{3eq34}, \ref{3eq61}), and each set can totally determine a solution for $\mathbf{U}_0$. For the period-1 solutions, we find that the width of PTR is always given by $\bar{Z}_{2,r}^{(1)}$  in (\ref{eq99}). The analysis  of $\bar{Z}_{2,r}^{(1)}$, including the dependence of material parameters, can be done easily. Many features are similar to those in the preceding section around (\ref{3eq51}). The main differences are that it slightly varies with $\gamma^\ast$ and is slightly larger than the width of PTR in two-region solutions, the $\bar{Z}_0^{(1)}$ in (\ref{3eq47}).

Since determination of $\bar{Z}_0$ from (\ref{3eq34}, \ref{3eq61}) is very sensitive to $\gamma^\ast$,  one could easily miss some solutions for a given $\gamma^\ast$, thereby missing part of stress-elongation curve. We find out that although one $\gamma^\ast$ corresponds to many solutions, one $\bar{Z}_0$ corresponds to only one period-1 solution. Therefore, from the perspective of numerical stability, it is a good strategy to control $\bar{Z}_0$ instead of $\gamma^\ast$ to find the solution set $(\gamma^\ast, \bar{Z}_2, \bar{Z}_0)$ by (\ref{3eq34}, \ref{3eq61}). In addition, the quantity  $\bar{Z}_0$ has a one-to-one correspondence to the total elongation $\Delta$ of cylinder, making it easier to compare with curves in displacement-controlled experiments. This strategy is adopted in the computation of solutions of $\mathbf{U}_0$ and stress-elongation (stress-strain) curves. For two different $\bar{Z}_0$, Figure \ref{3fig12} shows the profiles of the axial strain $W_{0Z}$ in a region near $Z=0.5$. It shows that the middle smooth curve in PTR connects the low strain austenite phase and high-strain martensite phase.  As $\bar{Z}_0$ increases (from Figure \ref{3fig12a} to Figure \ref{3fig12b}), the PTR or transformation front gradually moves from middle to the ends.

\begin{figure}
\subfigure[]{\includegraphics[width=2.5in]{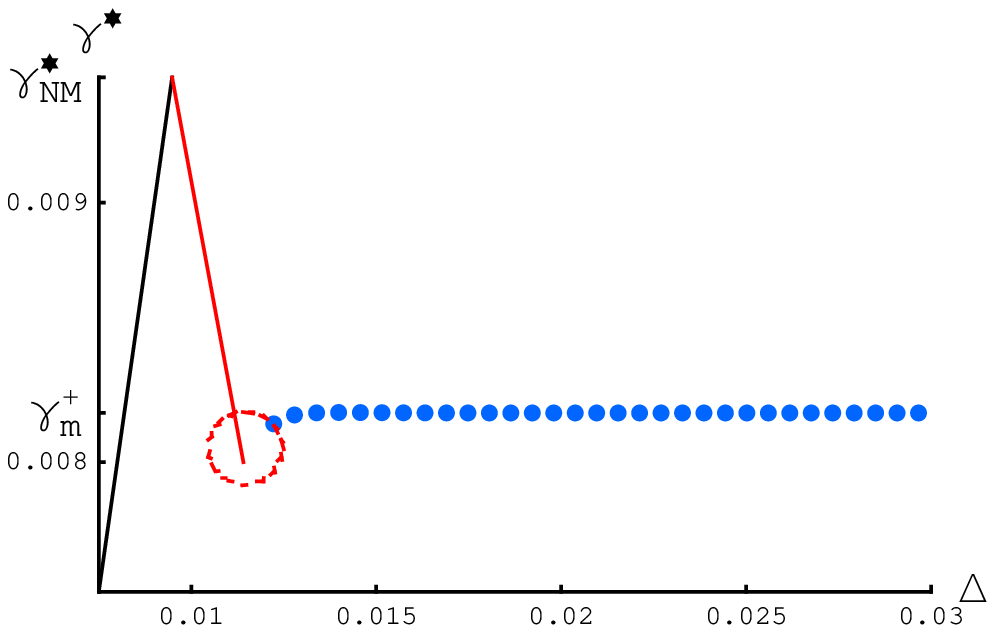}\label{3fig13a}}
\subfigure[]{\includegraphics[width=2.5in]{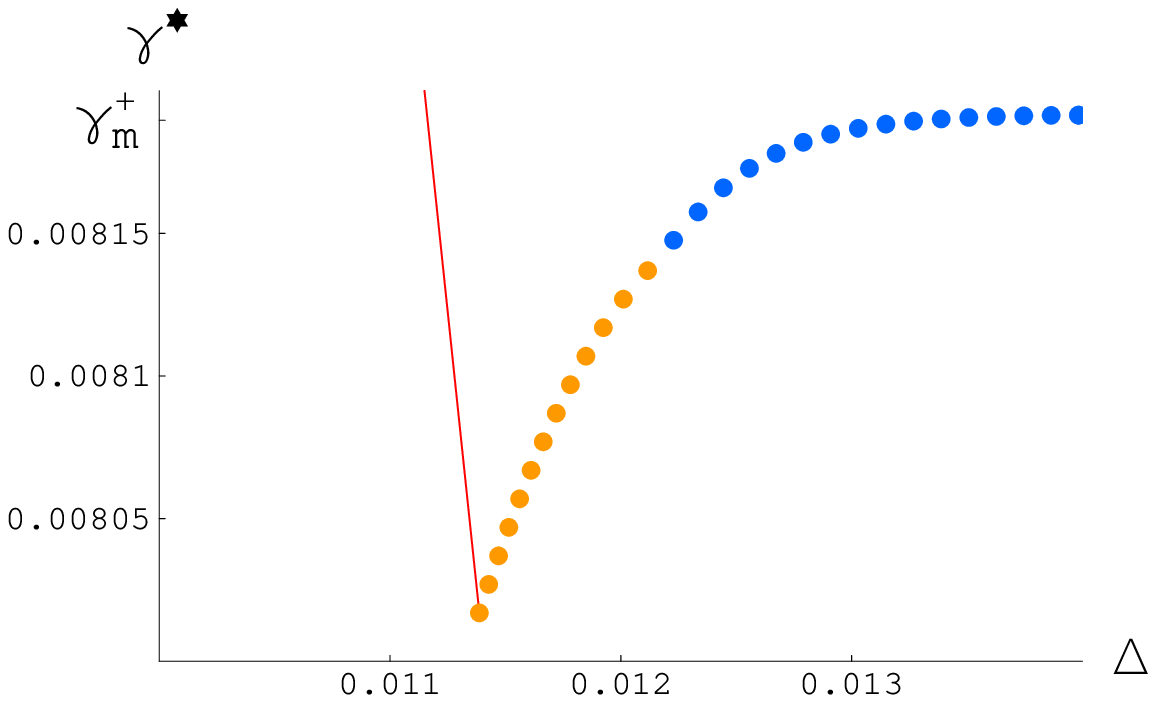}\label{3fig13b}}
\caption{ (a) The total elongation of period-1 three-region solutions with $a=0.03$ in blue dots, where the red line is from two-region solutions. (b) The enlarged part of the gap and the elongation of connecting solutions in yellow dots.}\label{3fig13}
\end{figure}

From the solution of the axial strain $W_{0Z}$, one can calculate the total
elongation of the cylinder by
\begin{equation}
\begin{aligned}
\label{3eq54_1} \Delta & = \int_0^{1} W_{0Z} (Z) \mathrm{dZ}\\
& = 2 \int_0^{Z_0} W_{0Z}^A (Z) \mathrm{dZ} + 2 \int_{Z_0}^{Z_2} W_{0Z}^P
(Z) \mathrm{dZ} +  2 \int_{Z_2}^{1/2} W_{0Z}^M
(Z) \mathrm{dZ}.
\end{aligned}
\end{equation}
With material constants in (\ref{eq75_1}) and $a=0.03$, Figure \ref{3fig13a} shows the stress-elongation curve of the period-1 three-region solutions, where each blue dot is from one solution and the reference red line is from the previous two-region solutions. The stress always stays very close to the Maxwell stress $\gamma_m^+$, which explains the sensitivity to $\gamma^\ast$. This feature is in good agreement with the stress plateau in the curves from many experiments \citep{tobushi,shaw1,sun2}.   As $\Delta$ decreases to about 0.012, the width of MR gradually decreases to 0. The three-region solutions then degenerates to one solution with essentially two regions at a limit case. However, the stress-elongation curves of two-region and three-region solutions are not smoothly connected (see the gap indicated by the red circle in Figure \ref{3fig13a}), so the limit three-region solution is not included in the previous two-region solutions. We will examine the connection between the two-region and three-region solutions in the next section.

\section{Connecting solutions}
The mismatch of the period-1 two-region and three-region solutions is due to
the interaction of interface $I_{PM}$ with the middle surface, and some intermediate solutions are missing. For the two-region solutions, the limit case is that some point on the middle surface $Z=0.5$ is on phase interface $\alpha=1$. More precisely, the center point of the middle surface first satisfies $\alpha=1$, when $\gamma^\ast= \gamma_c^\ast$ in (\ref{3eq53_1}), while other points do not. For the three-region solutions, the limit case
is that the MR disappears, i.e. the whole middle surface is exactly the interface $\alpha=1$ (they coincide). In between, there are solutions that the interface condition $\alpha=1$ is satisfied  from one
point of the middle surface to the whole surface. In this case, both the interface conditions and symmetric conditions should be taken into account for determining such intermediate solutions. Now we adopt one interface condition and one
symmetric condition 
\begin{equation}
\begin{aligned}
\label{3eq63} \alpha(0,0.5)=1,\quad \tilde{V}_{0Z}(0.5)=0,
\end{aligned}
\end{equation}
which are satisfied by both limit solutions. More explicitly, the condition $(\ref{3eq63})_1$ is given by
\begin{equation}
\begin{aligned}
\label{3eq64} &C_6 e^{-d_3 \bar{Z}_0} \bar{m}_6 + C_5 e^{d_3
\bar{Z}_0} \bar{m}_6+ C_7 \bar{m}_7 \cos
(d_4 \bar{Z}_0)+C_8 \bar{m}_7 \sin (d_4 \bar{Z}_0) \\
&= \bar{m}_8 \left( \gamma^\ast s_2- D_{\phi }^+ +D_k^+ \right),
\end{aligned}
\end{equation}
where the material constants $\bar{m}_6,..,\bar{m}_8$ are given in Appendix B.

Following similar analysis as in Section 4, the expressions of $C_1,..,C_4$ will
be the same (see $(\ref{3eq11})$ and above), but $C_5,..,C_8$ will be determined in terms of $\bar{Z}_0$ and $\gamma^\ast$ by $(\ref{eq87})_{2,3}$ and $(\ref{3eq63})$, whose explicit expressions are given in Appendix B. Finally the condition $(\ref{eq87})_1$  provides a single nonlinear algebraic equation for $\bar{Z}_0$
\begin{equation}
\begin{aligned}
\label{3eq67} f_3(\bar{Z}_0;\gamma^\ast)  =0,
\end{aligned}
\end{equation}
where the lengthy expression is omitted. For this connecting case, the above equation makes sense only for $\gamma^\ast$ between two critical stresses determined by the two
limit solutions in two-region and three-region cases. This small interval of $\gamma^\ast$ is below but not far from $\gamma^+_m$, and is independent of the radius $a$. For a given $\gamma^\ast$ in this small interval, there is a unique period-1 solution for
$\bar{Z}_0$, between the two values for the width of PTR in the limit
solutions.  This in turn determines all other unknowns and the period-1 solution for $\mathbf{U}_0$. At the two critical stresses, one can recover the two limit solutions of the previous sections, thus such intermediate solutions naturally connect the two-region and
three-region solutions.  With $a=0.03$ and material constants in (\ref{eq75_1}), Figure \ref{3fig13b} shows the total elongation of the above connecting solutions indicated by the yellow dots, which smoothly connect the red curve and blue points for the two-region and three-region solutions.


\section{The transition process}
\label{sec7}

Based on results in the previous sections, we now investigate the phase transition process for the displacement-controlled loading process (unloading process is similar). The optimal solution is chosen by the energy criteria, if multiple solutions exist. We adopt the total pseudo-elastic energy as
\begin{equation}
\begin{aligned}
\label{eq76} &\mathcal{W}^+ =\int_{\Omega}
[\Phi(\mathbf{F},\alpha)+\Phi_D^+(\alpha) ] \mathrm{d}\Omega,\qquad
\Phi_D^+ (\alpha)=\int_0^\alpha A^{+}(\alpha)\mathrm{d}\alpha.
\end{aligned}
\end{equation}
We focus on the homogeneous deformations and the inhomogeneous deformations from period-1 solutions in the previous sections, since one can easily show that period-k solution ($k\ge 2$) with larger width of PTR are not favored in  terms of the above energy  \cite{songdai2015}.

For the case of a relatively large radius $a>a_c = 1/K_2$, where $K_2$ is given in (\ref{3eq55})) and (\ref{3eq56}), the transition process is smooth and occurs gradually. Take $a=0.03$ with material constants in (\ref{eq75_1}) for example, Figure \ref{3fig13} shows that there is a unique period-1 solution  for each fixed total elongation $\Delta$ near nucleation of martensite phase. Therefore, the curve in displacement-controlled process goes exactly along the elongation curves in Figure \ref{3fig13}, with a gradual smooth transition from homogeneous deformations to localized inhomogeneous deformations. At the critical stress or elongation $\gamma^\ast = \Delta= \gamma^\ast_{NM}$, the homogenous deformation in austenite phase is replaced by the two-region solutions with fixed PTR, and the stress drops gradually as $\Delta$ further increases. Subsequently, the two region solutions are replaced by connecting solutions and three-region solutions, as the PTR moves along the cylinder as $\Delta$ increases.  The stress during propagation of PTR always stays close to Maxwell stress $\gamma_m^+$ as $\Delta$ increases. These features agree well with the experiments in \cite{tobushi,sun2}.

\begin{figure}
\begin{center}
\subfigure[]{\includegraphics[width=2.5in]{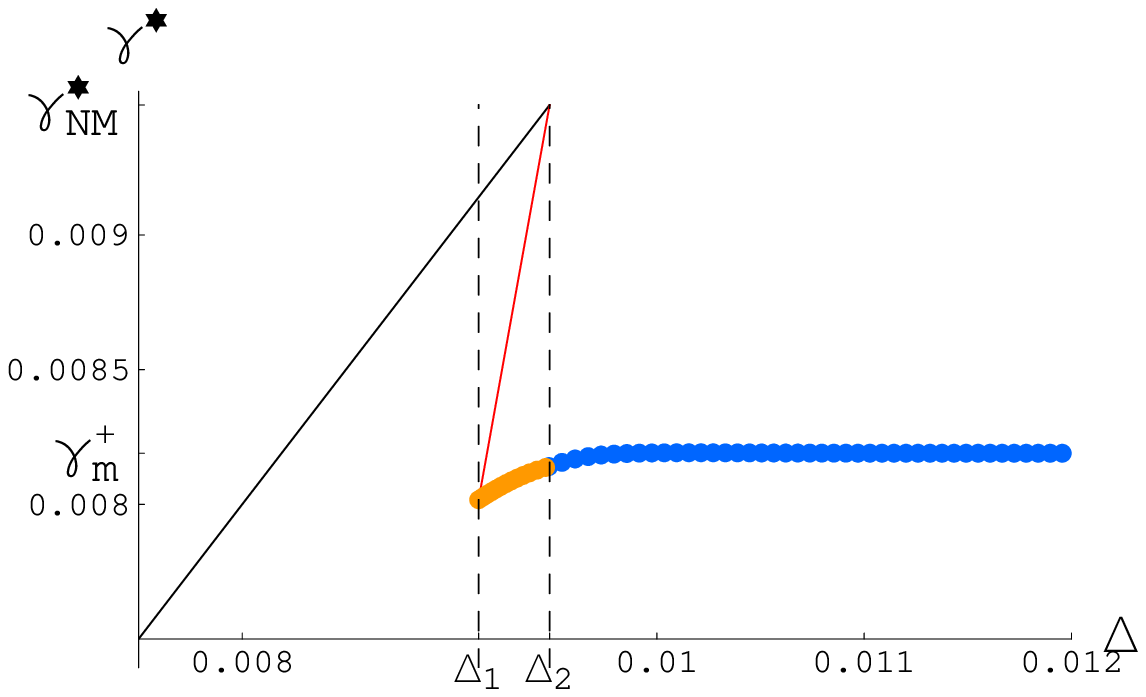}\label{3fig14a}}
\subfigure[]{\includegraphics[width=2.5in]{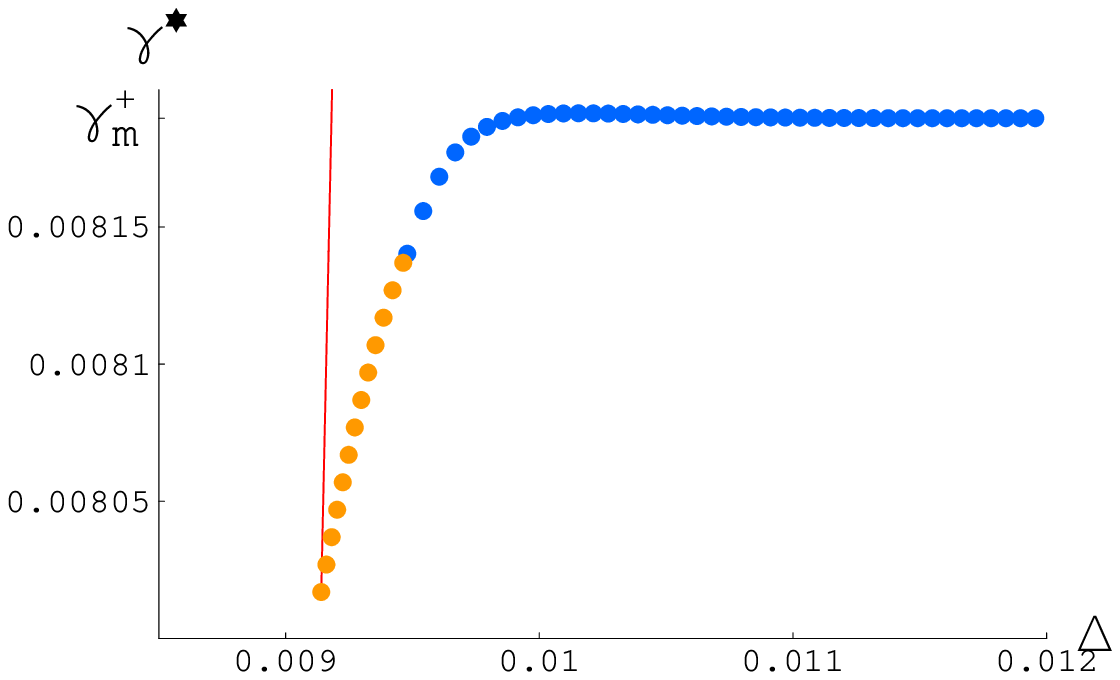}\label{3fig14b}}
\end{center}
\caption{ (a) The total elongation of homogeneous deformation and all period-1 inhomogeneous deformations near nucleation of martensite phase with $a=0.01$. (b)
The enlarged part of elongation of connecting
solutions in yellow dots.}\label{3fig14}
\end{figure}

\begin{figure}
\subfigure[]{\includegraphics[width=2.5in]{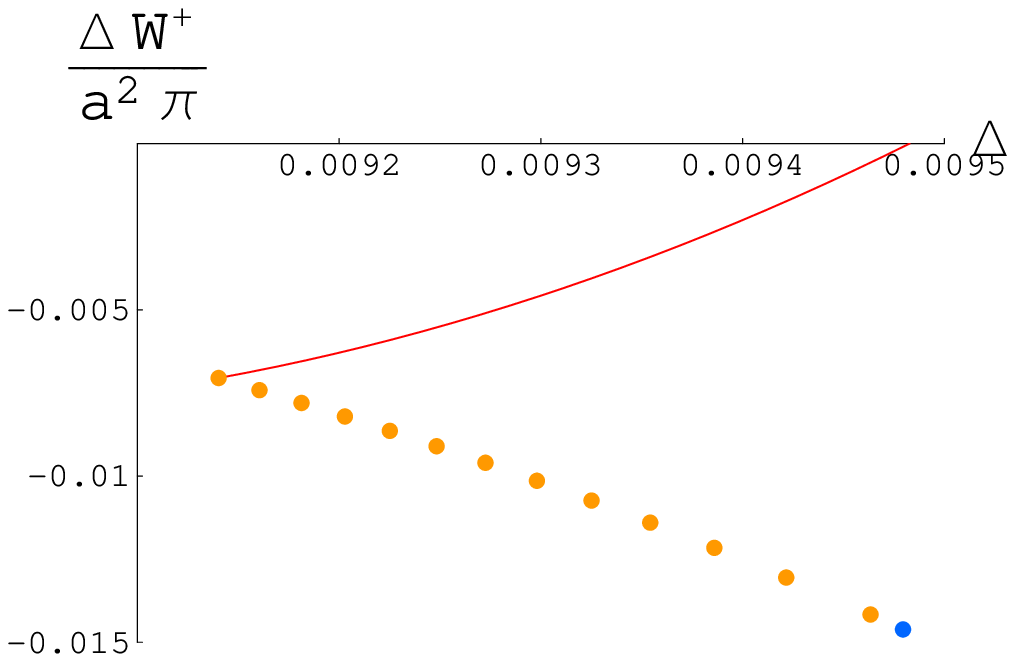}\label{3fig15a}}
\subfigure[]{\includegraphics[width=2.5in]{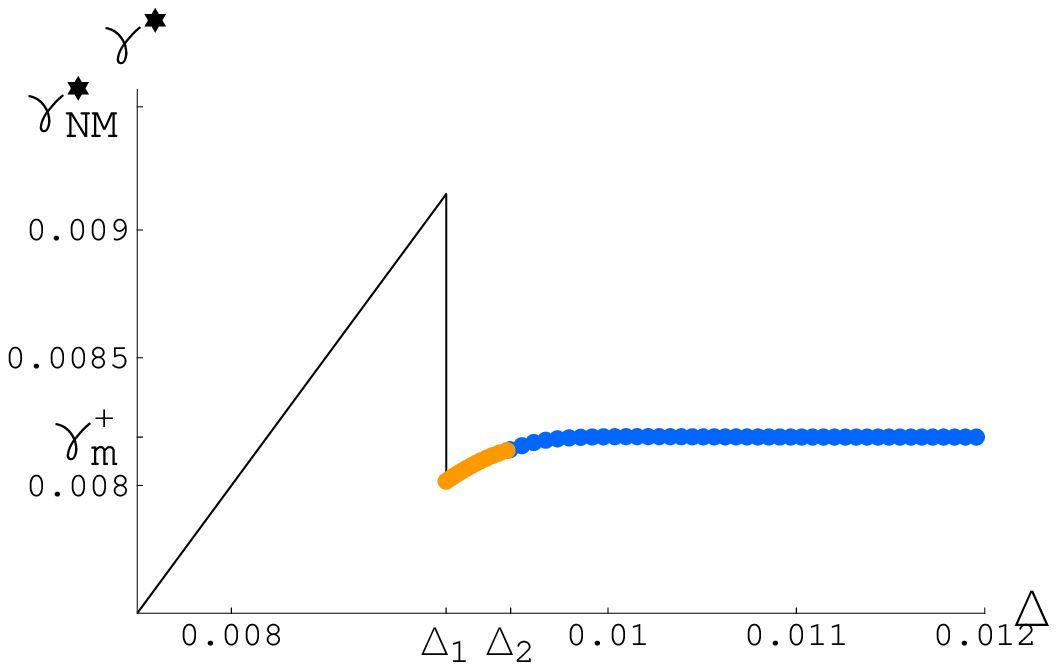}\label{3fig15b}}
\caption{ (a) Energy difference $\Delta \mathcal{W}^+$ in (\ref{eq_energy}) with $a=0.01$ (red curve for two region solutions, and dots for connecting and three-region solutions) (b) The stress-elongation curve with $a=0.01$ based on minimal energy criterion.}\label{3fig15}
\end{figure}

For the case of a relatively small radius $a<a_c$, the transition occurs abruptly, indicated by a jump in solutions in a snap-back bifurcation\footnote{One can also analogically think of the two cases with large $a$ and small $a$ as the supercritical and subcritical bifurcations used in dynamical systems.}. Take $a=0.01$ with material constants in (\ref{eq75_1}) for example, Figure \ref{3fig14} shows total elongation of homogeneous deformation in austenite phase and all the period-1 inhomogeneous deformations near the nucleation of martensite phase. It is clear that there exist three solutions for certain fixed elongation in $[\Delta_1,\Delta_2]$, where
\begin{equation}
\begin{aligned}
\label{3eq68} \Delta_1 \approx K_1 (\gamma_m^+ -
\gamma^\ast_{NM}) +\gamma^\ast_{NM} ,\quad\Delta_2= \gamma_{NM}^\ast,
\end{aligned}
\end{equation}
and $K_1$ is given in $(\ref{3eq55})$. As the radius $a$ decreases, $\Delta_1$ will decrease and hence the interval $[\Delta_1,\Delta_2]$ will become larger. The nucleation point lies within this interval, and will be determined by using the energy (\ref{eq76}). The energy difference is defined by
\begin{equation}
\label{eq_energy}
\begin{aligned}
\Delta \mathcal{W}^+ (sol_{inhomo})=  \mathcal{W}^+ (sol_{inhomo}) - \mathcal{W}^+ (sol_{homo}),
\end{aligned}
\end{equation}
where $\mathcal{W}^+ (sol_{homo})$ is the energy with the homogeneous deformation (solution) in austenite phase which is set as a reference value, and $\mathcal{W}^+ (sol_{inhomo})$ is the energy with any other inhomogeneous deformations. If $\Delta \mathcal{W}^+$ is positive, the homogeneous deformation is more stable. Otherwise if $\Delta \mathcal{W}^+$ is negative, the inhomogeneous deformations are more stable, and the one with minimal $\Delta \mathcal{W}^+$ is the optimal one. Figure \ref{3fig15a} shows the energy differences in (\ref{eq_energy}) scaled by a factor $\pi a^2$ for the period-1 inhomogeneous deformations. The red curve in Figure \ref{3fig15a} depicts the energy difference for the two-region solutions corresponding to the red line in Figure \ref{3fig14a}), and the dots depict the energy difference for the connecting solutions and the three-region solutions corresponding to dots in Figure \ref{3fig14a}. By the minimal energy criterion, the solution will jump from homogeneous state in austenite phase to a localized inhomogeneous state (the limit two-region solution) at $\Delta= \Delta_1$, as shown in Figure \ref{3fig15b}. Then the stress goes up a bit with connecting solutions and soon goes along the stress plateau with three-region solutions, corresponding to the propagation of  transformation front in experiments \cite{tobushi,sun2}.

The above quantitative analyses for large and small radius $a$ agree with the experimental observations in Figure 8 of \cite{sun2}. With relatively large $a$, the transition process in experiments is smooth and stress-strain curve is in agreement with Figure \ref{3fig13}. With relatively small radius of wire, the stress drop in experiments is very sharp, indicating an abrupt transition at one point, which is consistent with the sharp stress drop in Figure \ref{3fig15b}. This is due to the transition of solutions in a snap-back bifurcation in Figure \ref{3fig14a}. Figures 8 and 9 of \cite{sun2} also show decreasing peak stress with increasing radius $a$, which does not agree with above analysis. One reason could be that the above analysis is the ideal case by minimal energy criterion, whereas in practical case the local stability, the imperfection of the material and the perturbations during experiments could influence the exact nucleation (peak) stress (see remark below). The other reason could be that the current work neglects some high-order nonlinear terms in hyperelastic energy, the importance of which will be left for future investigation. 

\Remark From the above analysis, for a relatively small radius $a$ the stress will drop to the plateau at total elongation $\Delta_1$ before it reaches the peak of 1D intrinsic stress-strain curve at $\Delta_2$. This is the ideal case by minimal energy criterion, where the transition takes places at $\Delta_1$, by selecting the solution with global minimal energy.  In practical experiments, the transition point would be influenced by the energy difference (barrier) of the solutions, the imperfection of the material and perturbations in the experimental setup and operations.  As the homogeneous deformation is locally stable (local energy minimum) \citep{Ericksen}, the transition is very likely to occur after $\Delta_1$, until the perturbation is enough to overcome the energy barrier. For a relatively large $a$,  in ideal case transition takes place at $\Delta_2$. But since the energy barrier is 0 at this transition point, perturbations or imperfections can lead to a transition well before total elongation $\Delta$ reaches $\Delta_2$. Therefore, experimental curves with small $a$ tend to climb further along intrinsic linear curve at austenite phase in Figure \ref{fig2} and curves with large $a$ tend to concave down before reaching peak \cite{sun2}. As a result, experiments show higher nucleation stress (larger $\Delta$) for smaller $a$ in \cite{sun2}.

The unloading process can be investigated similarly. We briefly mention the main modifications. The phase transition criterion changes to $D= -A^{-}$, and the equation in PTR changes accordingly. The 1D homogeneous unloading curve and Maxwell stress $\gamma_m^{-}$ modify as in Figure \ref{fig2}. The boundary surfaces are initially in martensite phase, i.e., the boundary condition $(\ref{3eq5})_1$ should be used. The two-region solutions will contain MR and PTR,  and three-region solutions will have MR, PTR and AR from left to right in interval $[0,0.5]$.  The typical features of stress-elongation curve will be similar, except that there is stress rise  at nucleation of austenite phase instead of stress drop. 

\begin{figure}
\begin{center}
\subfigure[]{\includegraphics[width=2.5in]{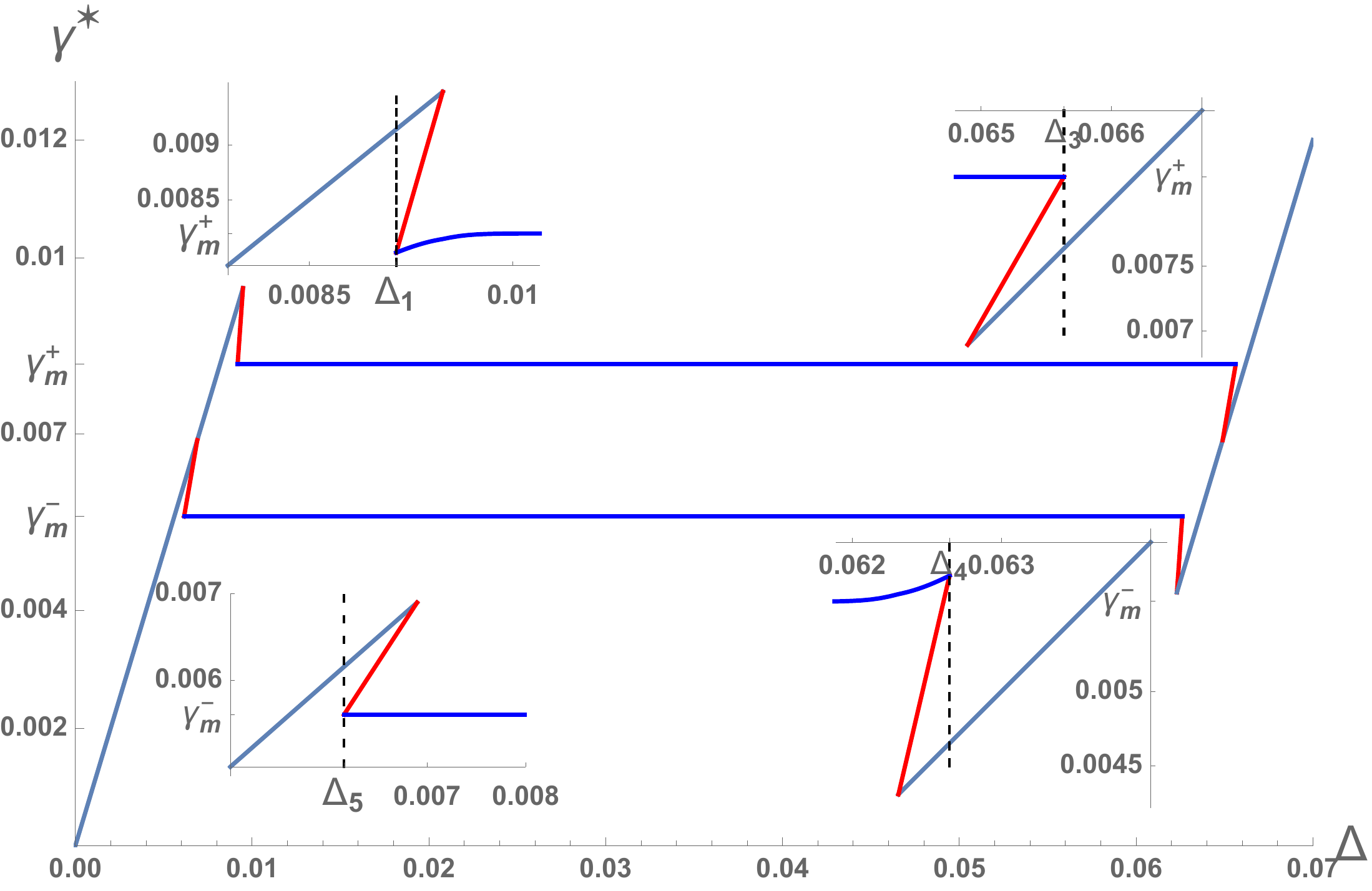}\label{3fig16a}}
\subfigure[]{\includegraphics[width=2.5in]{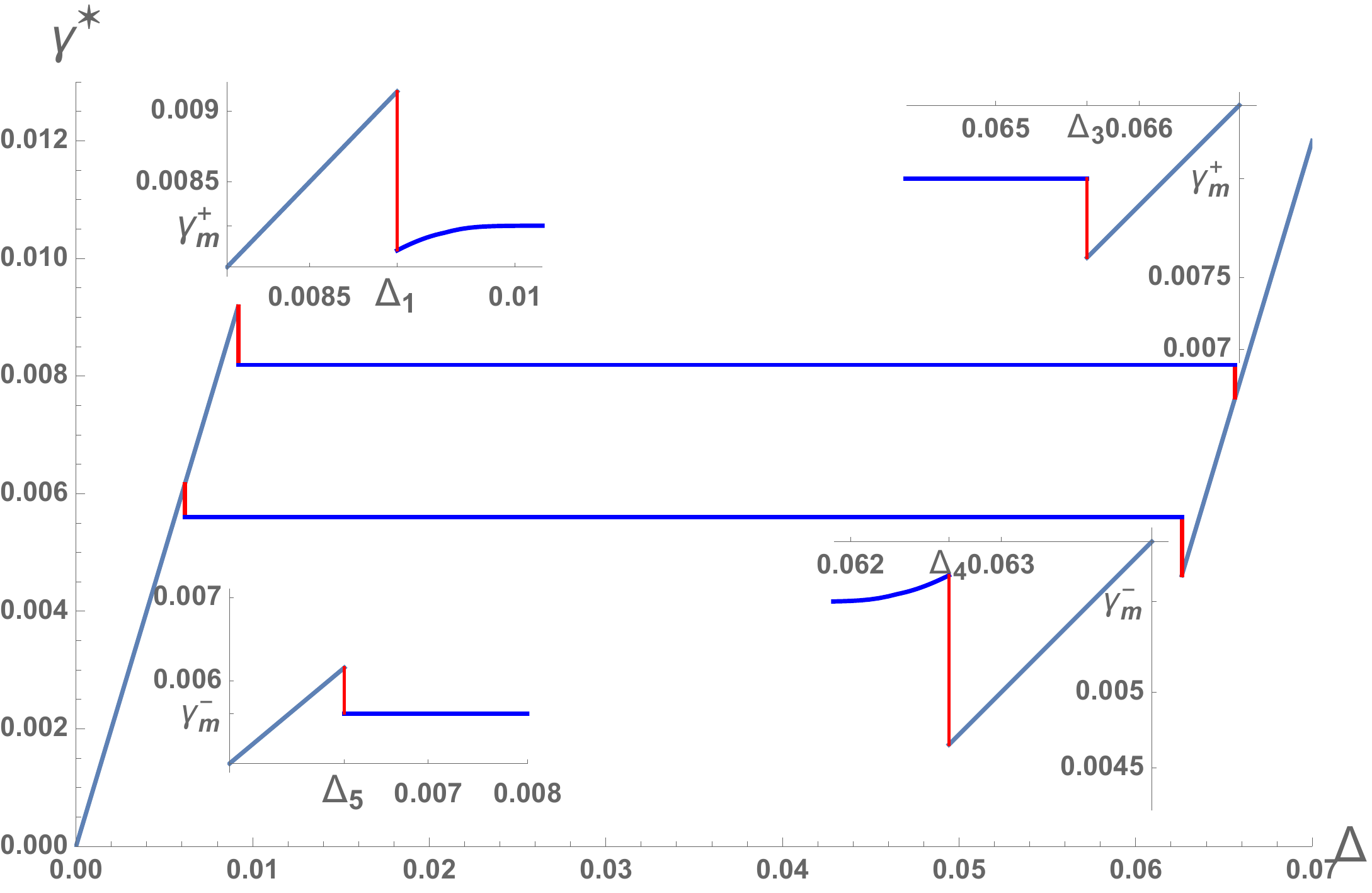}\label{3fig16b}}
\end{center}
\caption{ (a) The total elongations for homogeneous deformations and all period-1 inhomogeneous deformations with $a=0.01$. (b) The complete loading and unloading curves with $a=0.01$ by energy criterion.
}\label{3fig16}
\end{figure}

Finally, we present the stress-elongation curves for loading and unloading processes with $a=0.01$. Figure \ref{3fig16a} shows the total elongations for homogeneous deformations in austenite and martensite phases and inhomogeneous deformations from period-1 solutions.  Figure \ref{3fig16b} shows the complete loading and unloading curves based on energy criterion, which are in good agreement with experiments in \citep{tobushi,shaw1,sun2}. The four embedded figures are the enlarged parts of curves near nucleation and coalescence of austenite and martensite phases, where $\Delta_1,\Delta_3,\Delta_4,\Delta_5$ denotes the total elongations at the four transition points.

\section{Concluding remarks}

Based on a 3D constitutive model, we have systematically investigated  the phase transition process of SMA wire from homogeneous deformations to localized inhomogeneous deformations. The 3D system is simplified to a 1D piecewise linear system, while keeping essential high-dimensional effects. Free interfaces between different phase regions are included, and inhomogeneous deformations with two regions and three regions are studied analytically. The results clearly demonstrate important features in the transition process, such as the stress drop (rise) at nucleation of the martensite (austenite) phase and stress plateau at Maxwell stress during propagation of transformation front.  The roles of material and geometrical constants are revealed. The width of PTR is proportional to radius and the dependence on material parameters is mainly through a parameter associated with strain softening behaviour. As radius decreases, the stress drop at nucleation of the martensite phase tends to be very sharp, due to transition of solutions in a snap-back bifurcation. The snap-back bifurcation sheds lights on the difficulties of direct numerical computations, since it is not an easy task to capture all the possible solutions with free interfaces in simulation and select the optimal one from these solutions. This study could provide insights into the mechanism and instability of other materials or structures with strain softening behaviour, and shed light on designing functionally graded materials to avoid instability.

The present work is restricted to the symmetric case, where phase transition takes place in the middle. When phase transition occurs at one end, there will be complicated interaction between phase interface and end surface (like the connecting solution in this paper), which will be left for future investigation. We also plan to extend the work to the investigation of geometrically and functionally graded SMAs.

\section*{Acknowledgement}
The work was done at CityU and supported by a grant from the National Nature
Science Foundation of China (Project No.: 11572272) and GRF grant
(Project No.: CityU 11302417) from the Research Grants Council
of Hong Kong SAR, China.

\appendix

\section{Constants and expressions in section 2}

The two moduli $\xi_1$ and $\xi_2$ are the first-order incremental elastic moduli of the strain energy $\Phi_1$ or $\Phi_2$ (assumed the same at first order) \cite{songdai2015}, which are related to the lam\'{e} constants  \cite{ogden} (or Young's modulus $E$ and Poisson's ratio $\nu$)
\begin{equation}
\begin{aligned}
\lambda= \xi_2, \quad \mu=\frac{1}{2}(\xi_1 -\xi_2), \quad E=
\frac{\xi _1^2+\xi _2 \xi _1-2 \xi _2^2}{\xi _1+\xi _2}, \quad \nu =
\frac{\xi _2}{\xi _1+\xi _2}.
\end{aligned}
\end{equation}

The row vectors $\mathbf{b}_k^{(1)},\mathbf{b}^{(2)}, \bar{\mathbf{b}}_k^{(1)}, \bar{ \mathbf{b}} ^{(2)}$ ($k\ge 0$) are given by
\begin{equation}
\begin{aligned}
\mathbf{b}_k^{(1)}= &\left(\xi _2,\frac{(2 k+1) \xi _1+\xi _2}{2 (k+1)}\right), \quad \mathbf{b}^{(2)}=\left(-\xi _1,-\xi _2\right),\\
\bar{\mathbf{b}}_k^{(1)}=& [-\frac{(2 B-k^{+}) \xi _2-s_1 s_2 \left(\xi
_1^2+\xi _2 \xi _1-2 \xi _2^2\right)}{2 \left(\xi _1+\xi
_2\right) s_1^2-4 s_2 \xi _2 s_1-2 B+k^{+}+s_2^2 \xi _1},\\
& \frac{(2 k + 1)\xi _1+\xi _2}{2 (k+1)}-\frac{\left(s_2 \xi _2-s_1
\left(\xi _1+\xi _2\right)\right)^2}{2 \left(\xi _1+\xi _2\right)
s_1^2-4 s_2 \xi _2 s_1-2B+k^{+}+s_2^2 \xi _1}],\\
\bar{\mathbf{b}}^{(2)}=&[ -\frac{2 s_1^2 \xi _1^2+\left(2 \xi _2
s_1^2-2 B+k^{+}\right) \xi _1-4 s_1^2 \xi _2^2}{2 \left(\xi_1+\xi
_2\right) s_1^2-4 s_2 \xi _2 s_1-2 B+k^{+}+s_2^2 \xi _1},\\
& -\frac{(k^{+}-2 B) \xi _2+s_1 s_2 \left(\xi _1^2+\xi _2 \xi _1-2
\xi_2^2\right)}{2 \left(\xi _1+\xi _2\right) s_1^2-4 s_2 \xi _2 s_1-2
B+k^{+}+s_2^2 \xi _1}].
\end{aligned}
\end{equation}

The matrix $\mathbf{H}$ is given by
\begin{equation}
\begin{aligned}
\label{eq39} & \mathbf{H}= \mathbf{A}^{-1} \left(
\begin{array}{c}
\frac{1}{4} \mathbf{b}_1^{(1)} \\ \frac{1}{16} \mathbf{b}^{(2)} \\
\end{array} \right)^{-1} \left( \begin{array}{c}
\mathbf{b}_0^{(1)} \\ \frac{1}{2} \mathbf{b}^{(2)} \\
\end{array} \right),\quad  \mathbf{A}=\frac{1}{\xi_1 -\xi_2} \left(
\begin{array}{cc}-2 \xi _1 & - (\xi _1 + \xi _2) \\
\xi _1+\xi _2 & 2 \xi _2
\end{array}
\right).
\end{aligned}
\end{equation}
Similarly $\bar{\mathbf{H}}$ is defined in the same way by replacing $\mathbf{A},\mathbf{b}_0^{(1)},\mathbf{b}_1^{(1)}, \mathbf{b}^{(2)}$ with $\bar{\mathbf{A}},\bar{\mathbf{b}}_0^{(1)},\bar{\mathbf{b}}_1^{(1)}, \bar{\mathbf{b}}^{(2)}$, see also Appendix B of \cite{songdai2015}. The vectors $\mathbf{f}_0, \mathbf{f}_1,\bar{\mathbf{f}}$ are given by
\begin{equation}
\begin{aligned}
& \mathbf{f}_0= \mathbf{A}^{-1} \left(
\begin{array}{c} \frac{1}{4} \mathbf{b}_1^{(1)} \\
\frac{1}{16} \mathbf{b}^{(2)} \\
\end{array} \right)^{-1} \left( \begin{array}{c}
0 \\  -\frac{\gamma}{2} \\
\end{array}\right),\\
& \mathbf{f}_1= \mathbf{A}^{-1} \left(
\begin{array}{c} \frac{1}{4} \mathbf{b}_1^{(1)} \\
\frac{1}{16} \mathbf{b}^{(2)} \\
\end{array} \right)^{-1} \left( \begin{array}{c}
s_2 \xi_2 -s_1 \xi_1 -s_1 \xi_2 \\ -\frac{1}{2}[\gamma + s_2 \xi_1 -2 s_1 \xi_2] \\
\end{array}\right),\\
& \bar{\mathbf{f}}= \bar{\mathbf{A}}^{-1} \left(
\begin{array}{c} \frac{1}{4} \bar{\mathbf{b}}_1^{(1)} \\
\frac{1}{16} \bar{\mathbf{b}}^{(2)} \\
\end{array} \right)^{-1} \left( \begin{array}{c}
(s_2 \xi_2 -s_1 \xi_1 -s_1 \xi_2) \alpha_{00} \\ -\frac{1}{2}[\gamma + (s_2 \xi_1 -2 s_1 \xi_2)\alpha_{00}] \\
\end{array}\right),
\end{aligned}
\end{equation}
where for loading process $\alpha_{00}$ is 
\begin{equation}
\begin{aligned}
\alpha_{00}=&\frac{{B}+{Y}^{+}
-{\phi_1}+{\phi _2}}{-2 \xi _1 s_1^2-2 \xi _2 s_1^2+4 s_2 \xi _2 s_1+2
{B}-{k}^{+}-s_2^2 \xi _1}.
\end{aligned}
\end{equation}

\section{Constants and expressions in solutions}

The constants $d_1,d_2,q_1,q_2$ are given by
\begin{equation}
\begin{aligned}
&d_1=\frac{2}{a} \sqrt{\sqrt{g}+g}, \qquad d_2=\frac{2}{a}
\sqrt{\sqrt{g} -g},\qquad g(\nu):=\frac{\nu +1}{\nu +3},\\
& q_1=\frac{2\nu ^2-3 \nu-1}{2 \nu ^2+\nu +1},\quad q_2=
\frac{\sqrt{2} \sqrt{\nu +1} (2 \nu -1)}{2 \nu ^2+\nu +1}.
\end{aligned}
\end{equation}
The  quantities $d_3,d_4,\bar{q}_{1},\bar{q}_2$ can be expressed by the components of $\bar{\mathbf{H}}$, and with $\epsilon_1:= D_k^{\pm}/s_2^2 $, we get
\begin{equation}
\begin{aligned}
\label{eq57} & d_3\approx \frac{2 \sqrt{2} \sqrt[4]{5-4 \nu }
\sqrt[4]{D_k^{\pm}}}{a \sqrt{s_2}} + \frac{\sqrt{2}  (\nu +1)
(D_k^{\pm})^{3/4} }{a \sqrt[4]{5-4 \nu } s_2^{3/2}},\\
& d_4\approx \frac{2 \sqrt{2} \sqrt[4]{5-4 \nu } \sqrt[4]{D_k^{\pm}}}{a \sqrt{s_2}} -
\frac{\sqrt{2}  (\nu +1) (D_k^{\pm})^{3/4} }{a \sqrt[4]{5-4 \nu }
s_2^{3/2}},\\
& \bar{q}_{1}\approx -1 - \frac{3 (1 - 2 \nu )
\sqrt{D_k^{\pm}}}{\sqrt{5-4 \nu } s_2},\qquad
\bar{q}_{2}\approx -1 + \frac{3 (1 - 2 \nu ) \sqrt{D_k^{\pm}}}{\sqrt{5-4 \nu } s_2}.\\
\end{aligned}
\end{equation}

The constants in the analytical results are given by
\begin{equation}
\begin{aligned}
q_{11}=&\frac{ \left(\left(2 \nu ^2+\nu +1\right) s_1-\nu (\nu +3)
s_2\right)}{2 s_1 \left(\nu s_1-s_2\right)},\\
 q_{22}=& -
\frac{\sqrt{\nu +1} \left(\left(2 \nu ^2+\nu +1\right) s_1+(\nu +3)
s_2\right)}{2 \sqrt{2} s_1 \left(\nu s_1-s_2\right)},\\
\bar{q}_{11}=&[D_k^{\pm} \left(\left(4 \nu +\bar{q}_{1} \left(2 \nu
^2+\nu +1\right)\right) s_1-(\nu +3) ((\bar{q}_{1} -1) \nu +1) s_2\right)\\
&-2 s_1 \left(\nu s_1-s_2\right) \left(2 s_1+\bar{q}_{1}  s_2\right)]
/[2 (\bar{q}_{1} -\bar{q}_{2}  ) D_k^{\pm} s_1\left(\nu  s_1-s_2\right)],\\
\bar{q}_{22}=&[D_k^{\pm} \left(\left(4 \nu +\bar{q}_{2} \left(2 \nu
^2+\nu +1\right)\right)
s_1-(\nu +3) ((\bar{q}_{2}  -1) \nu +1) s_2\right)\\
&-2 s_1 \left(\nu s_1-s_2\right) \left(2 s_1+\bar{q}_{2} s_2\right)]
/[2 (\bar{q}_{1} -\bar{q}_{2}  ) D_k^{\pm} s_1\left(\nu
s_1-s_2\right)],\\
n_{11}= &d_4   \bar{q}_{2} \bar{q}_{11},\qquad n_{12}=- d_3 \bar{q}_{1}
\bar{q}_{22}, \\
 n_{13}= & q_1 d_1 q_{11} - q_2 d_2 q_{11} + q_2 d_1
q_{22} + q_1 d_2 q_{22},\\
\bar{m}_6= & (-\bar{q}_1 -2 \nu ) s_1+((\bar{q}_1 -1) \nu +1) s_2,\\
\bar{m}_7= & (-\bar{q}_2 -2 \nu ) s_1+((\bar{q}_2 -1) \nu +1) s_2,\\
\bar{m}_8= & [2 s_1^2-4 \nu  s_2 s_1-(\nu -1) s_2^2+\left(2 \nu ^2+\nu
-1\right) D_k^+ ]/D_k^+ .\\
\end{aligned}
\end{equation}

The integrating constants $C_9,..,C_{12}$ in three-region solutions are given by
\begin{equation}
\begin{aligned}
\label{3eq58} C_9= & -\bar{C} \left(e^{ d_1  (2  \bar{Z}_0 + \bar{Z}_2
)} \left( e^{2 d_1   \bar{Z}_2 }  q_{11}  \cos(d_2 (2 \bar{Z}_0 -
\bar{Z}_2 ))+
e^{2 d_1   \bar{Z}_0 }  q_{11}  \cos ( d_2 \bar{Z}_2 ) \right.\right. \\
&\left.\left. q_{22} \left(e^{2 d_1 \bar{Z}_2 } \sin ( d_2 (2
\bar{Z}_0 - \bar{Z}_2 ))+e^{2 d_1\bar{Z}_0 }
\sin ( d_2 \bar{Z}_2 )\right)\right)\right),\\
C_{10}= & \bar{C} \left( e^{ d_1  (2  \bar{Z}_0 + \bar{Z}_2 )}
\left(-q_{22} e^{2 d_1 \bar{Z}_2 } \cos ( d_2 (2 \bar{Z}_0 - \bar{Z}_2
)) -q_{22} e^{2  d_1   \bar{Z}_0 } \cos(d_2 \bar{Z}_2 )\right.\right. \\
&\left.\left. + q_{11}  \left(e^{2  d_1   \bar{Z}_2 } \sin ( d_2 (2
\bar{Z}_0 - \bar{Z}_2 ))+e^{2 d_1   \bar{Z}_0 } \sin ( d_2
\bar{Z}_2 )\right)\right)\right),\\
C_{11}= & -\bar{C}\left( e^{ d_1   \bar{Z}_2 } \left(  e^{2  d_1
\bar{Z}_0 } q_{11} \cos ( d_2  (2  \bar{Z}_0 - \bar{Z}_2 ))+ q_{11}
e^{2d_1 \bar{Z}_2 }  \cos ( d_2   \bar{Z}_2 ) \right.\right. \\
&\left.\left. -q_{22} \left(e^{2  d_1 \bar{Z}_0 } \sin ( d_2 (2
\bar{Z}_0 - \bar{Z}_2 ))+e^{2 d_1 \bar{Z}_2}
\sin ( d_2   \bar{Z}_2 )\right)\right)\right),\\
C_{12}= &\bar{C} \left( e^{ d_1   \bar{Z}_2 } \left(-q_{22} e^{2  d_1
\bar{Z}_0 } \cos ( d_2  (2  \bar{Z}_0 - \bar{Z}_2 ))
-q_{22} e^{2 d_1 \bar{Z}_2 } \cos ( d_2 \bar{Z}_2 ) \right.\right. \\
&\left.\left. - q_{11} \left(e^{2  d_1   \bar{Z}_0 } \sin ( d_2 (2
\bar{Z}_0 - \bar{Z}_2 ))+e^{2  d_1 \bar{Z}_2 } \sin ( d_2 \bar{Z}_2
)\right)\right)\right),\\
\bar{C}= & \frac{\left(D_k^+ -D_{\phi }^+ +\gamma^\ast
s_2\right)}{\left(\left(2 e^{2 d_1 ( \bar{Z}_0 + \bar{Z}_2 )} \cos (2
d_2 ( \bar{Z}_0 - \bar{Z}_2 ))+e^{4  d_1 \bar{Z}_0 }+e^{4  d_1
\bar{Z}_2 }\right)\right)}.
\end{aligned}
\end{equation}

The integrating constants $C_5,..,C_8$ in connecting solution are given by
\begin{equation}{\fontsize{11pt}{\baselineskip}
\begin{aligned}
\label{3eq66} C_5= &\tilde{C} \left(d_4 e^{d_3 \bar{Z}_0} \bar{m}_8
\bar{q}_2 \cos (d_4 \bar{Z}_0) D_k^+ -\left(d_4 \bar{q}_2 \left(e^{d_3
\bar{Z}_0} \bar{m}_8-\bar{m}_6 \bar{q}_{22}\right) \cos (d_4
\bar{Z}_0)\right.\right.\\
&\left.\left.+\bar{m}_7 \left(d_4 e^{d_3 \bar{Z}_0} \bar{q}_2
\bar{q}_{11}-d_3 \bar{q}_1 \bar{q}_{22} \sin (d_4
\bar{Z}_0)\right)\right) \left(D_{\phi }^+ -\gamma^\ast s_2\right)\right),\\
C_6= &\tilde{C} \left(e^{d_3 \bar{Z}_0} \left(\left(d_4 \bar{q}_2
\left(\bar{m}_8-e^{d_3 \bar{Z}_0} \bar{m}_6 \bar{q}_{22}\right) \cos
(d_4 \bar{Z}_0)\right.\right.\right.\\
&\left.+\bar{m}_7 \left(d_4 \bar{q}_2 \bar{q}_{11}+d_3 e^{d_3
\bar{Z}_0} \bar{q}_1 \bar{q}_{22} \sin (d_4 \bar{Z}_0)\right)\right)
\left(D_{\phi }^+ -\gamma^\ast
s_2\right)\\
&\left.\left. -d_4 \bar{m}_8 \bar{q}_2 \cos (d_4
\bar{Z}_0) D_k^+ \right)\right),\\
C_7= &\bar{q}_{11} \left(D_{\phi }^+ -\gamma^\ast  s_2\right),\\
C_8= &-\tilde{C} \left(d_3 \left(1+e^{2 d_3 \bar{Z}_0}\right)
\bar{m}_8 \bar{q}_1 D_k^+ -\left(d_3 \bar{q}_1 \left(e^{2 d_3
\bar{Z}_0} \bar{m}_8+\bar{m}_8-2 e^{d_3 \bar{Z}_0} \bar{m}_6
\bar{q}_{22}\right)\right.\right.\\
&+d_3 \left(1+e^{2 d_3 \bar{Z}_0}\right) \bar{m}_7
\bar{q}_1 \bar{q}_{11} \cos (d_4 \bar{Z}_0) \\
&\left.\left. +d_4 \left(-1+e^{2 d_3 \bar{Z}_0}\right) \bar{m}_6
\bar{q}_2 \bar{q}_{11} \sin (d_4
\bar{Z}_0)\right) \left(D_{\phi }^+ -\gamma^\ast  s_2\right)\right),\\
\tilde{C}=&1/[d_4 \left(-1+e^{2 d_3 \bar{Z}_0}\right) \bar{m}_6
\bar{q}_2 \cos (d_4 \bar{Z}_0)-d_3 \left(1+e^{2 d_3 \bar{Z}_0}\right)
\bar{m}_7 \bar{q}_1 \sin (d_4 \bar{Z}_0)].
\end{aligned}}
\end{equation}

\section{Material constants and one numerical solution}

The following material constants are used to show numerical results and figures
 \begin{equation}
\label{eq75_1}
\begin{aligned}
&B/E=k^{\pm}/E=1.5\times 10^{-4},\quad Y^{\pm}=0, \quad \Delta\phi/E=4
\times 10^{-4},\\
& s_1=0.03,\quad s_2=0.058,\quad E=4*10^4 ~\mathrm{MPa}, \quad
\nu=1/3.
\end{aligned}
\end{equation}
One two-region solution from whole system by Newton's method is given by
\begin{equation}
\begin{aligned}
&C_1=0,\quad C_2=0, \quad C_3=0.000262346,\quad C_4=0.00289176,\\
& C_5=-1.29457*10^{-6}, \quad
C_6=-0.00354165,\quad C_7=-0.012628,\\
& C_8=-0.00173087,\quad \bar{Z}_0=0.0570056.
\end{aligned}
\end{equation}


\section*{References}
\bibliographystyle{model1-num-names}
\bibliography{reference}


%
%
%
\end{document}